\documentclass[lettersize,journal]{IEEEtran}

\usepackage{textcomp}
\usepackage{stfloats}
\usepackage{url}
\usepackage{verbatim}
\usepackage{color,array,amsthm}
\usepackage{bm}
\usepackage{amsfonts}
\usepackage{graphicx}
\usepackage[justification=centering]{caption}
\usepackage{amsmath}
\usepackage{booktabs}
\usepackage{siunitx}
\usepackage{amssymb}
\usepackage{eucal}
\usepackage{mathrsfs}
\usepackage{multirow}
\usepackage{cite}
\usepackage{stfloats}
\usepackage{threeparttable}
\usepackage{subfig}
\usepackage{lineno,hyperref}
\usepackage{makecell}
\usepackage{algpseudocode}
\usepackage{amsmath,amsthm}
\newtheorem{theorem}{Theorem}
\usepackage[ruled,linesnumbered]{algorithm2e}
\usepackage{algpseudocode}
\usepackage{makecell}
\usepackage{xcolor}
\usepackage{threeparttable}

\begin{document}

\title{Class Information Guided Reconstruction for Automatic Modulation Open-Set Recognition}

\author{Ziwei Zhang,~\IEEEmembership{Student Member,~IEEE}, Mengtao Zhu ~\IEEEmembership{Member,~IEEE}, Jiabin Liu ~\IEEEmembership{Member,~IEEE}, \\Yunjie Li,~\IEEEmembership{Senior Member,~IEEE}, Shafei Wang

\thanks{This research was supported by National Natural Science Foundation of China (NSFC) under grants nos.62306036 {\itshape (Corresponding author: Jiabin Liu)}.}

\thanks{Ziwei Zhang is with School of Cyberspace Science and Technology, Beijing Institute of Technology, Beijing, 100081, China, (e-mail:{3120225659@bit.edu.cn}). Jiabin Liu is with School of Information and Electronics, Beijing Institute of Technology, Beijing, 100081, China, (e-mail:{liujiabin@bit.edu.cn}). Mengtao Zhu is with School of Cyberspace Science and Technology, Beijing Institute of Technology, Beijing, 100081, China (e-mail: {zhumengtao@bit.edu.cn}). Yunjie Li is with School of Information and Electronics, Beijing Institute of Technology, Beijing, 100081, China, (e-mail:{liyunjie@bit.edu.cn}). Shafei Wang is with School of Cyberspace Science and Technology, Beijing Institute of Technology, Beijing, 100081, China, and also with the Laboratory of Electromagnetic Space Cognition and Intelligent Control, Beijing, 100191, China. }
}

\markboth{Journal of \LaTeX\ Class Files,~Vol.~14, No.~8, August~2021}%
{Shell \MakeLowercase{\textit{et al.}}: A Sample Article Using IEEEtran.cls for IEEE Journals}


\maketitle

\begin{abstract}
Automatic Modulation Recognition (AMR) is vital for radar and communication systems.  Traditional AMR operates under a closed-set scenario where all modulation types are pre-defined. However, in practical settings, unknown modulation types may emerge due to technological advancements. Closed-set training poses the risk of misclassifying unknown modulations into existing known classes, leading to serious implications for situation awareness and threat assessment. To tackle this challenge, this paper presents a Class Information guided Reconstruction (CIR) framework that can simultaneously achieve Known Class Classification (KCC) and Unknown Class Identification (UCI). The CIR leverages reconstruction losses to differentiate between known and unknown classes, utilizing Class Conditional Vectors (CCVs) and a Mutual Information (MI) loss function to fully exploit class information. The CCVs offer class-specific guidance for reconstruction process, ensuring accurate reconstruction for known samples while producing subpar results for unknown ones. Moreover, to enhance distinguishability, an MI loss function is introduced to capture class-discriminative semantics in latent space, enabling closer alignment with CCVs during reconstruction. The synergistic relationship between CCVs and MI facilitates optimal UCI performance without compromising KCC accuracy. The CIR is evaluated on simulated and real-world datasets, demonstrating its effectiveness and robustness, particularly in low SNR and high unknown class prevalence scenarios.

\end{abstract}


\section{Introduction}
\label{intro}


\par Automatic Modulation Recognition (AMR) pertains to the classification of intra-pulse modulation types, which has a wide range of applications in military and civilian fields \cite{TCCN2,TVT4, RN605, RN603}. Within military applications, AMR aids the inference of radar intentions, facilitates effective radar operation and contributes to further specific emitter identification (SEI) \cite{TCCN1, RN501, RN504}. For civilian use, AMR provides modulation details necessary for spectrum management \cite{TCCN3, RN413, RN503} and monitors the spectrum to prevent malicious attacks. These capabilities of AMR make it an essential technology for enabling dynamic spectrum access and ensuring physical-layer security for Internet of Things (IoT) networks \cite{RN601}.

\begin{figure}[!t]
    \centering
    \includegraphics[width=3.5in]{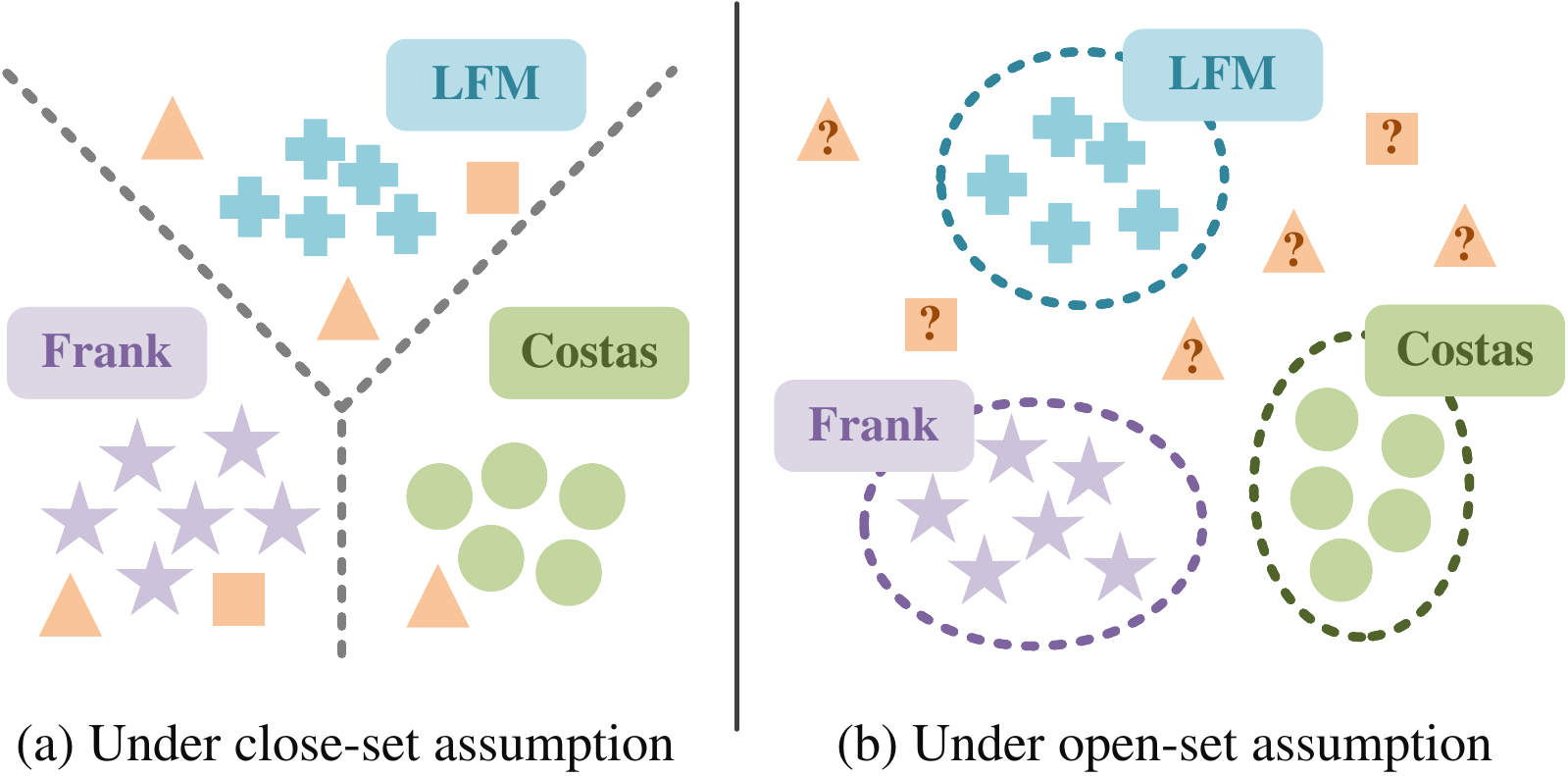}
    \captionsetup{justification=justified}
    \caption{Comparison of closed-set and open-set assumptions. Under close-set assumption, samples from unknown classes are forcibly misclassified to one of known classes. Under open-set assumption, the unknown modulation types are assigned to a distinct category denoted as the unknown class.} 
    \label{fig0}
\end{figure}

\par As the development of Deep Learning (DL), many DL-based AMR approaches have demonstrated promising performance\cite{RN402, TCCN4,TCCN5,RN403,RN405,RN406,RN407,RN505, RN602, RN606} under a closed-set assumption, wherein the modulation types in the training and test sets are identical. However, the closed-set paradigm poses limitations for real-world electromagnetic environments, wherein unknown modulation types not seen during training would be encountered \cite{RN408} due to evolving wireless technologies, interference sources, environmental changes, etc. Under a closed-set assumption, such unknown samples would be forcibly assigned to one of the known classes, as shown in Fig. \ref{fig0} (a). This mis-classification can lead to inaccurate situation awareness by perceiving unknown threats as familiar ones. Additionally, defensive techniques may no longer be optimally matched to adversary systems whose types were not precisely determined.


\par To address the issue of mis-classification for unknown samples, Automatic Modulation Open-Set Recognition (AMOSR) is introduced to handle with modulation types that are not encountered during training \cite{RN101,RN115,RN116,RN506}. Specifically, when encountering unknown modulation types when testing, we explicitly assign them to a distinct category denoted as the unknown class, instead of erroneously labeling them as an existing class, as presented in Fig. \ref{fig0} (b). The AMOSR formulates two objectives: Known Class Classification (KCC) and Unknown Class Identification (UCI), which is better aligned with real-world electromagnetic environments. However, existing AMOSR approaches face two key challenges in accomplishing these objectives: (1) Inappropriate decision boundary; (2) Sparse feature distribution.




\par The first challenge in AMOSR is inappropriate decision boundary, which usually stems from incompatible assumptions. Specifically, certain methods relied on the assumption that certain samples were naturally distorted versions of their classes \cite{RN111,RN112,RN113} and could be utilized to delineate the decision boundary. However, when extending this assumption directly to AMOSR, a significant number of samples at low SNRs would be incorrectly labeled as unknown. This is because distortions in signal samples primarily arise from noise contamination, rather than fundamentally altering the modulation-specific time-frequency signatures. 

\par The second challenge in AMOSR involves addressing sparse feature distributions exacerbated by noise. Conventional Open-Set Recognition (OSR) techniques aimed to concentrate latent features within single class while increasing margins between classes to attain better OSR performance \cite{RN102,RN213,RN214}. However, declining SNR poses unique difficulties for AMOSR. Specifically, at lower SNRs, within-class features become more dispersed as features between classes start to overlap, leading to significant performance degradation. While previous studies investigated OSR tasks on degraded images, such as blurry or noisy photographs, these approaches prove to be ineffective for AMOSR. This is primarily due to the significant corruption and distortion present in Time-Frequency Images (TFI) generated from noise-contaminated received signals.

\par To tackle the aforementioned challenges, this paper proposes a Class Information guided Reconstruction (CIR) framework for AMOSR. The CIR distinguishes known and unknown classes based on reconstruction losses, which provides a reliable metric unaffected by sample deformities. By fully mining class information to guide the reconstruction process, CIR is able to amplify the distance between discriminative features (i.e. reconstruction losses) of known and unknown classes, which would alleviate the adverse impacts incurred by decision boundary selection. The CIR leverages class information by introducing Class Conditional Vectors (CCVs) and Mutual Information (MI) loss function. By the designed CCVs, known samples matching to a CCV can be perfectly reconstructed, whereas unknown samples not corresponding to any CCVs yield larger reconstruction losses. Concurrently, maximizing the MI between classes and latent representations encourages the extraction of class-focused representations, which allows for better matching to CCVs during reconstruction. Meanwhile, minimizing the MI between samples and representations helps disentangle the semantics from input variations, enhancing generalization performance. Furthermore, a denoising module is introduced to mitigate the adverse effects of noise, providing clearer information for subsequent reconstruction stages and ensuring the robustness of CIR at low SNRs.

\par The main contributions can be summarized as follows:

\begin{itemize}

 \item This paper proposes a CIR framework that pioneeringly employs reconstruction losses to tackle the AMOSR task, providing a robust solution for the open-set challenges in real-world electromagnetic environment.

 \item The elaborately designed CCVs embed class information into the latent feature space to guide the reconstruction process, amplifying the discrepancy of reconstruction losses between known and unknown samples, thereby enhancing the UCI performance.
 
 \item We introduce an MI loss function to further excavate class information, which makes latent features better aligned with the CCVs to ensure the discriminability of unknown classes. Concurrently, this function lessens the influence of sample distribution, thereby strengthening the generalization ability of our method.

 \item Comprehensive experiments on simulated and real signals demonstrate CIR outperforms state-of-the-art OSR baselines, especially under challenging conditions such as low SNRs and high proportions of unknown classes.

    
\end{itemize}


\par This paper is organized as follows. Section \ref{section2}  reviews related OSR methods in Computer Vision (CV) and AMR domains. Section \ref{section3} establishes preliminaries, including signal models, Time-Frequency Analysis (TFA) methods, and problem definition of AMOSR. In Section \ref{section4}, the proposed CIR is introduced, while the experimental design and the analysis of the results are presented in Section \ref{section simu}. Finally, the paper is concluded and future work is discussed in Section \ref{section6}.

\par The notations used in this paper are summarized in Table \ref{label_note}.

\begin{table}[htbp]
  \centering
  \caption{Important Notations}
    \begin{tabular}{p{6em}p{16em}}
    \toprule
    Notations & Definitions \\
    \midrule
    $\bm{\psi}$     & Modulation type \\
    $M$     & Number of modulation types \\
    $M_u$     & Number of unknown modulation types \\
    $N$     & Number of samples \\
    $\mathcal{D}$     & Dataset \\
    $\mathcal{L}$     & Loss function \\
    $\mathcal{H}$    & Fully connected layer\\
    $x_n$     & Raw TFI \\
    $x$     & Denoised TFI \\
    $y$     & Class label \\
    $\bm{z}$     & Latent representations \\
    $H$     & Height of the TFI \\
    $W$     & Width of the TFI \\
    $C$     & Number of channels \\
    $\bm{F}$     & Features in Restormer \\
    $\bm{R}$     & Residual image in Restormer \\
    $l$     & Class conditional vector \\
    $f$     & Frequency \\
    $\gamma$ & Lagrange multiplier \\
    $\tau$   & Decision threshold \\
    \bottomrule
    \end{tabular}%
  \label{label_note}%
\end{table}%

\section{Related Work}
\label{section2}
\subsection{OSR Methods in CV Domain}
\emph{Discriminative-based approaches}: By leveraging the powerful representation learning capability of deep networks, some classical discriminative-based approaches have emerged for OSR task. A plain choice was to utilize the maximum SoftMax probabilities and rejected low-confidence predictions~\cite{RN212}. Bendale et al.~\cite{RN105} proved SoftMax probability was not robust and proposed to replace the SoftMax function with the OpenMax function, which redistributed the scores of SoftMax to obtain the confident score of the unknown class explicitly. In~\cite{RN111}, Schlachter et al. proposed to split given data into typical and atypical normal subsets, thus finding a precise boundary to distinguish known classes from unknowns. Yang et al.~\cite{RN213} proposed generalized convolutional prototype learning, which replaced the close-world assumed SoftMax classifier with an open-world oriented prototype model. Chen et al.~\cite{RN102} developed Reciprocal Point Learning (RPL), which performed UCI task based on the otherness with reciprocal points. Subsequently, RPL was further improved to ARPL~\cite{RN214}, integrating an extra adversarial training strategy to enhance the model distinguishability by generating confusing training samples.

\emph{Generative-based approaches}: A number of studies have explored generative-based approaches, wherein Auto-Encoder (AE)-based and reconstruction-based techniques are mainly related to our work. Huang et al.~\cite{RN301} proposed a class-specific semantic reconstruction method, that replaced prototype points with manifolds represented by class-specific AEs and measured class belongingness through reconstruction error. In~\cite{RN104}, a Conditional Gaussian Distribution Learning (CGDL) was designed, which utilized Variational Auto-Encoder (VAE) to learn class conditional posterior distributions for OSR tasks. Yoshihashi et al.~\cite{RN103} designed the Classification-Reconstruction learning for OSR (CROSR), which used latent representations for closed set classifier training and unknown detection. In \cite{RN303}, a structured Gaussian mixture VAE was proposed, guaranteeing separable class distributions with known variances in its latent space. Sun et al.~\cite{RN304} designed an AE-based method that learned feature representations by modeling them as mixtures of exponential power distributions. In \cite{RN305}, a VAE contrastive learning method was proposed combined with Pseudo Auxiliary searching strategy to discover a more robust structure.


\par The preceding discriminative-based and generative-based approaches face significant difficulties as mentioned in Section \ref{intro}. Discriminative-based methods encounter challenges in establishing appropriate decision boundaries in high-dimensional space, leading to the excessive rejection of signals with low SNRs as unknowns. Meanwhile, generative-based approaches struggle to aggregate features, thereby failing to achieve sufficiently promising performance.


\subsection{OSR Methods in AMR Domain}

\emph{Machine learning-based approaches}: Chakravarthy et al.~\cite{RN207} designed a quantile one-class support vector machine-based algorithm. In \cite{RN208}, isolation forest models were used for each known signal class to perform detection of possible unknown signals. Han et al.~\cite{RN202} proposed a deep class probability output network and gave the confidence of identification results through the Kolmogorov–Smirnov test. While Chen et al.~\cite{RN204} established probability models of known classes based on extreme value theory. In \cite{RN206}, an OSR technology based on Dempster–Shafer evidence theory was proposed. 

\emph{Deep learning-based approaches}: Xu et al.~\cite{RN112} proposed a hybrid OSR approach combining modified Intra-Class Spitting (ICS) method with adversarial samples generation. Subsequently, he ~\cite{RN113} enhanced this via Transformer, attaining more precise boundaries. In \cite{RN201}, an algorithm leveraging vision Transformer, Wasserstein distance and reciprocal points improved OSR efficiency. Gong et al.~\cite{RN209} developed an improved counterfactual GAN architecture with multi-tasking mechanism for signal representation. In \cite{RN504}, Dong et al. proposed a zero-shot learning framework, where signal recognition and reconstruction neural networks were designed to tackle the OSR problem. An OSR model based on transfer deep learning and linear weight decision fusion was designed in \cite{RN203}. Zhang et al.~\cite{RN205} proposed a modified generalized end-to-end loss to increase the similarity of the same modulation type and reduce the similarity of different types. 




\par The above-mentioned AMOSR methods still face the following problems: (1) some studies directly applied CV-inspired frameworks without adequately addressing key differences in distortion mechanisms between visual and signal data; (2) while some existed work made some improvement by incorporating modulation characteristics, many still relied on relatively simplistic OSR techniques such as ML-based or OpenMax-oriented methods. In a word, further exploration is still needed for AMOSR to address the aforementioned issues. 

\section{Preliminaries}
\label{section3}
\subsection{Signal Model}
\par From a receiver’s perspective, signals contaminated with noise can be expressed as:

\begin{equation}
    \label{eq1}
    s[k]=x[k]+n[k],0 \leq k \leq K
\end{equation}
where $x[k]$ represents the noise-free signal, $n[k]$ represents the noise generated due to temperature of transmitter equipment, the receiver part of antenna, etc., $k$ is the sample index, and $K$ is the number of sampling points. 

\par The mathematical expression of $x[k]$ is given as:

\begin{equation}
    \label{eq3}
    x[k]=a[k]e^{j\theta[k]}
\end{equation}
where $a[k]$ is the non-zero constant envelope (i.e. the amplitude) within the pulse interval $\tau_{pw}$, and usually $a[k]=A$ for $0\leq kT_{s}\leq \tau_{pw}$, where $T_s$ is the time interval corresponding to the sampling frequency $f_s$. The instantaneous phase $\theta[k]$ can be defined through the instantaneous frequency $f[k]$ and the phase offset $\phi[k]$ as: 

\begin{equation}
    \label{eq4}
    \theta[k]=2\pi f[k](kT_s)+\phi [k]
\end{equation}


\subsection{Time Frequency Analysis}

\par Time-Frequency Analysis is employed to analyze signal variations in both the time and frequency domains. After TFA, the resulting TFIs can be fed into subsequent networks for feature extraction. The Short-Time Fourier Transform (STFT) is one commonly used technique for TFA.

\par The STFT involves dividing the signal into multiple short windows and applying the Fourier Transform to each window. By sliding these windows along the time axis, a comprehensive time-frequency representation of the entire signal is obtained as follows 

\begin{equation}
    \text{STFT}(t,f) = \int_{\infty}^{\infty}x(\delta)h(\delta-t)e^{-j2\pi f\delta} \text{d}\delta
\end{equation}
where $h(\delta-t)$ is the window function with the length of $\delta$, $f$ is the frequency.

\subsection{Problem Definition of AMOSR}

\begin{figure*}[!t]
    \centering
    \includegraphics[width=7.25in]{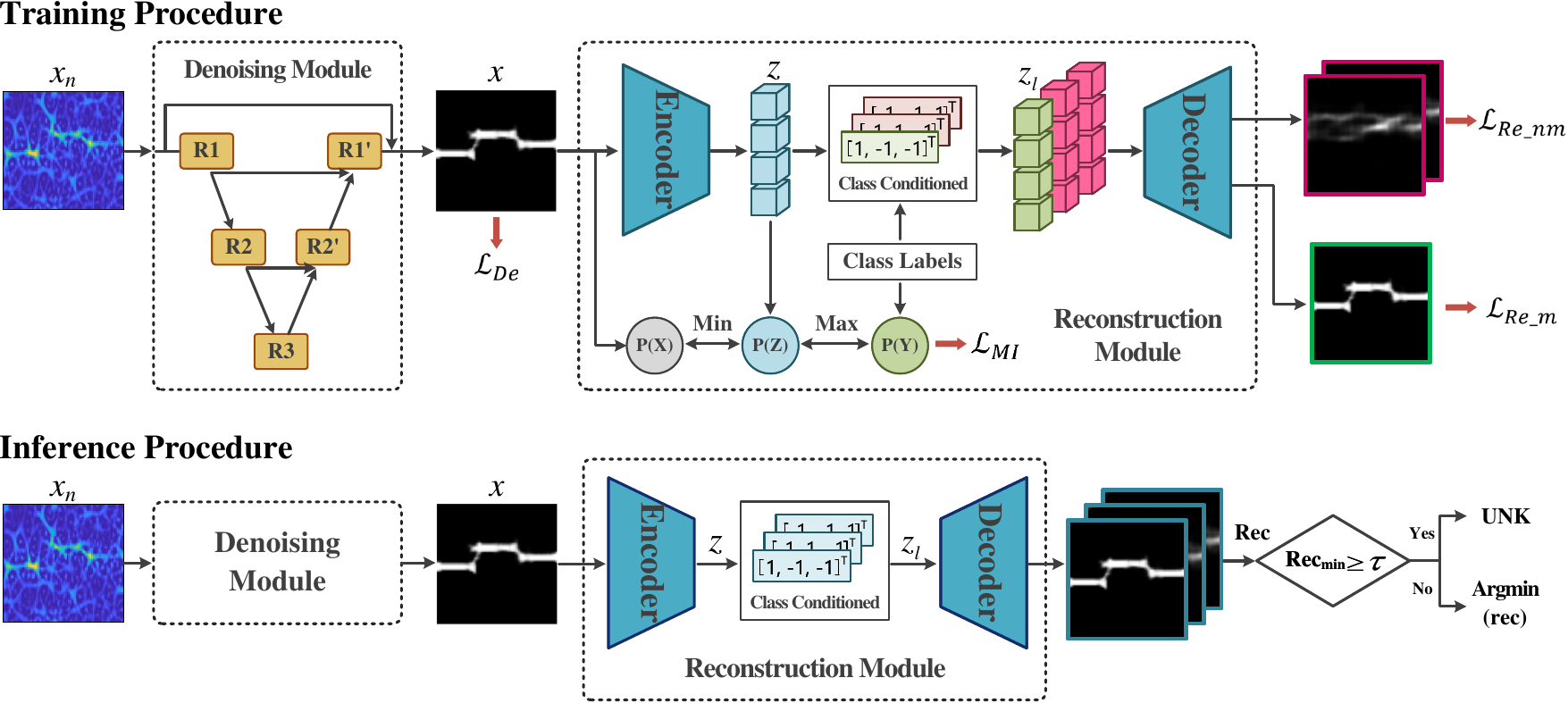}
    \captionsetup{justification=justified}
    \caption{Class information guided reconstruction framework. The proposed CIR consists of two core modules: a denoising module and a reconstruction module. The denoising module receives the raw TFI $x_n$ and obtains the denoised TFI $x$. The reconstruction module employs an AE as the backbone, where we propose CCVs and MI loss function to refine the latent representations $\bm{z}$ with class information, effectively increasing the distinguishability. In the inference procedure, the reconstruction losses are utilized to discriminate between known and unknown classes.}
    \label{fig2}
\end{figure*}

\par Considering $M$ known modulation types, the set of class labels can be denoted as $\bm{\Psi}=\{\psi_1,\psi_2,\cdots,\psi_M\}$. Given a labeled Training datasets $\mathcal{D}_{l}=\{\bm{x}^i_{l},y^i_{l}\}_{i=1}^{N_l}$ where labels $y^i_{l}\in\bm{\Psi}$, and a test set $\mathcal{D}_{t}=\{\bm{x}^i_{t},y^i_{t}\}_{i=1}^{N_{t}}$ where the label $y^i_{t}$ belongs to $\bm{\Psi} \cup \{\psi_{M+1},\psi_{M+2},\cdots,\psi_{M+M_u}\}$, where $M_u$ is the number of unknown classes that would be encountered in realistic scenarios. AMOSR can be divided into two sub-tasks: KCC and UCI. The goal of KCC is to learn a model from $\mathcal{D}_{\mathcal{L}}$ that can accurately predict class labels for test samples whose true labels are contained within $\bm{\Psi}$. The goal of UCI is to have the model assign a unified unknown label $\psi_{M+1}$ to samples whose true labels belong to $\{\psi_{M+1},\psi_{M+2},\cdots,\psi_{M+M_u}\}$.

\section{Method}
\label{section4}

\subsection{Overall Framework}
\par The proposed CIR utilizes reconstruction losses to distinguish known and unknown classes, which consists of two core modules: a denoising module and a reconstruction module, as presented in Fig. \ref{fig2}. The denoising module is introduced before reconstruction to ensure its functionality at low SNRs. Restormer is utilized as the backbone to denoise the raw input TFI $x_n$ and obtain noise-reduced signal $x$. The reconstruction module employs an AE structure to reconstruct the denoised signal $x$. In this module, the CCVs and MI loss function are elaborately designed to mine the class-information, thus enlarging the reconstruction discrepancy between known-class and unknown-class samples. In the training procedure, only latent representations $\bm{z}$ extracted from $x$ that match the CCVs can achieve perfect reconstruction, and MI loss function ensures reliable matching.

\par Since unknown-class samples cannot match any CCVs, their reconstruction losses will be significantly larger than the reconstruction loss for any known-class sample. Therefore, the inference procedure utilizes the difference in reconstruction losses to identify potential unknown classes. Inputs that result in high reconstruction losses can be flagged as an unknown class, while those identified as known classes will be output with their certain labels.

\subsection{Denoising Module}

\par Noise is a crucial factor influencing the performance in AMOSR. As the SNR decreases, recognition accuracy severely deteriorates. To improve AMOSR performance under low SNR conditions, the denoising module removes the noise in raw TFIs to make the discriminative information available at subsequent reconstruction stage. This facilitates more concentrated feature distributions within each class, ensuring the functionality of CIR at low SNRs.
\par In this paper, Restormer~\cite{RN108} is utilized to realize the denoising module owing to its favorable properties of high efficiency and promising performance. Restormer's symmetric encoder-decoder structure and lightweight components allow it to remove noise with much higher efficiency compared to heavy CNN-based models, suitable for real-time applications. By leveraging multi-scale feature propagation through skip connections, Restormer can effectively recover fine texture details lost to noise at both low and high resolutions within TFIs. By mitigating noise interference, Restormer furnishes the CIR framework with important robustness under low SNR conditions close to practical scenarios.

\par
\begin{figure}[!t]
    \centering
    \includegraphics[width=3.2in]{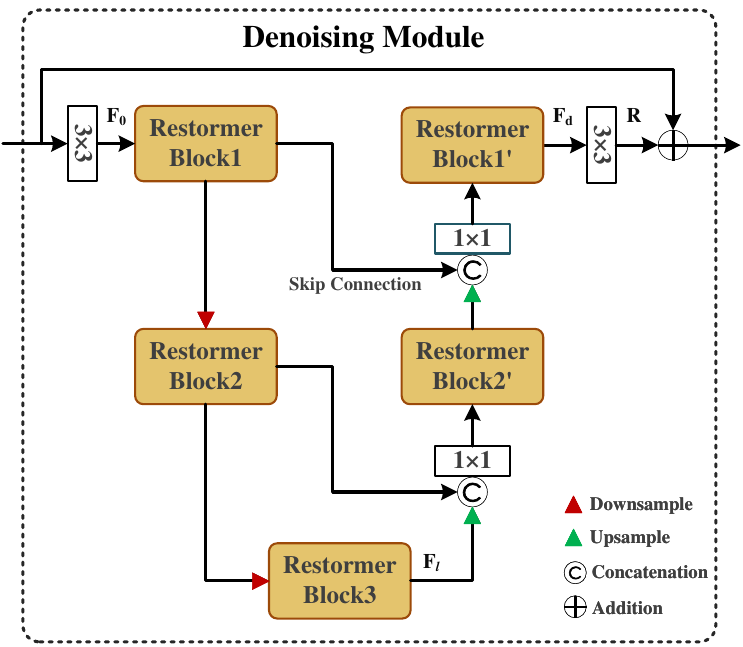}
    \captionsetup{justification=justified}
    \caption{Restormer-based denoising module.}
    \label{fig2_11}
\end{figure}

 The overall pipeline of the Restormer is presented in Fig. \ref{fig2_11}. Given a raw input TFI $x_n \in {\mathbb{R}}^{H\times W \times 1}$, Restormer first applies a 3x3 convolution to extract initial feature maps $\bm{F}_0 \in \mathbb{R}^{H\times W \times C}$, where $H\times W$ denotes the spatial dimension and $C$ is the number of channels. These shallow features $\bm{F}_0$ are then passed through a 2-level symmetric encoder-decoder and transformed into deep features $\bm{F}_d \in \mathbb{R}^{H\times W \times 2C}$. Each level of encoder-decoder contains multiple Restormer blocks, where the number of blocks is gradually increased from the top to bottom levels to maintain efficiency. The encoder hierarchically downsmaples spatially while expanding channels, extracting latent features $\bm{F}_l \in \mathbb{R}^{\frac{H}{4}\times \frac{W}{4} \times 4C}$. The decoder takes low-resolution latent features $\bm{F}_l$ as input and progressively recovers the high-resolution representations. Pixel unshuffling and shuffling are used for downsampling and upsampling respectively~\cite{RN109}. To assist the recovery process, the encoder features are concatenated with the decoder features via skip connections~\cite{RN110}. Channel reduction is applied via 1x1 convolution at each level. Finally, a convolution layer generates residual image $\bm{R}\in \mathbb{R}^{H\times W\times 1}$, which is added to the degraded input to obtain the restored image: $\bm{x} = \bm{x}_n + \bm{R}$.

\par The Restormer block is composed of two core components: (a) multi-Dconv head transposed attention (MDTA) and (b) gated-Dconv feed-forward network (GDFN). These components enable Restormer to achieve highly efficient computation while maintaining strong modeling capabilities. Details of the Restormer block design can be found in~\cite{RN108}. For brevity, the technical details are omitted here as they have been well described in the literature. 

\par The loss function for the denoising module is derived by calculating the Mean Squared Error (MSE) between the denoised TFIs and the ground truth noise-free TFIs as follows:

\begin{equation}
    \label{loss_de}
    \mathcal{L}_{De}=||\bm{x}_n-\bm{x}||_2^2
\end{equation}


\subsection{Reconstruction Module}
\par The reconstruction module introduces an AE structure to reconstruct the denoised signal $x$. Intuitively, known classes would be reconstructed more accurately with lower reconstruction losses compared to unknown classes. However, to enhance the UCI performance, we want the reconstruction loss distributions between known and unknown classes to be more divergent. Since only known class data is available during training, we maximize the utilization of known class information to guide the reconstruction process.

\par In this study, we elaborately design CCVs and MI loss function to refine the latent representations with class information, thus increasing the distinguishability for UCI task. The CCVs can match the latent representations, achieving perfect reconstruction for known samples and worse reconstruction for unknown samples. Meanwhile, the MI loss function aims to increase the relevance between the latent representations and class labels, while reducing the dependence on the input samples, which ensures reliable matching and improves generalization performance.


\par As presented in Fig. \ref{fig2}, the encoder maps the input $\bm{x}$ to low-dimensional latent representations $\bm{z}$. Then, $\bm{z}$ is combined with the designed CCVs to generate the class-conditioned representations $\{\bm{z}_1,\bm{z}_2,\cdots,\bm{z}_M\}$. Finally, the decoder maps $\{\bm{z}_1,\bm{z}_2,\cdots,\bm{z}_M\}$ back to the original data space $\{\hat{\bm{x}}_1,\hat{\bm{x}}_2,\cdots,\hat{\bm{x}}_M\}$, named class-conditioned reconstruction. Meanwhile, the proposed CIR is trained to learn disentangled latent representations by minimizing the MI between the latent representations $\bm{z}$ and the input data $\bm{x}$, while maximizing the MI between the $\bm{z}$ and the corresponding class labels $y$.

\subsubsection{Class-conditioned Reconstruction}

\par The CCVs are deliberately designed to provide class-relevant guidance for the reconstruction procedure. Specifically, each CCV aims to encapsulate the salient attributes of its corresponding class. By matching the latent representations to the CCVs during reconstruction, the AE is implicitly encouraged to embed class-discriminative information into the representations. This can help the AE perfectly reconstruct the known-class samples and poorly reconstruct the unknown-class samples, thereby enhancing the discrepancy in reconstruction losses between known-class and unknown-class samples.





\par Regarding the latent representations $\bm{z}$ obtained from the encoder, the CCV $l_j$ is defined as

\begin{equation}
      l_j(k)=
                \left\{
                        \begin{aligned}
                        &+1,k=j\\
                        &-1,k\neq j
                        \end{aligned}
                \right.  \quad  k,j\in \{1,2,\cdots,N\}
\end{equation}

\par Then the class-conditioned latent representations $\{\bm{z}_1,\bm{z}_2,\cdots,\bm{z}_N\}$ are obtained by the following equations,

\begin{equation}
    \bm{\alpha}_j = \mathcal{H}_{\bm{\alpha}}(l_j), \quad \bm{\beta}_j = \mathcal{H}_{\bm{\beta}}(l_j)
\end{equation}

\begin{equation}
    \bm{z}_j = \bm{\alpha}_j \odot \bm{z} +\bm{\beta}_j
\end{equation}

\par Here, $\mathcal{H}_{\bm{\alpha}}$ and $\mathcal{H}_{\bm{\beta}}$ are fully connected layers, $\bm{\alpha}_j$ and $\bm{\beta}_j$ have the same shape as vector $\bm{z}$, $\odot$ represents the Hadamard product.

\par The decoder is expected to perfectly reconstruct the original input when the CCV matches the class identity of the input, referred as the match condition (${\cdot}_m$). The perfectly reconstruction is achieved by minimizing the MSE of the reconstructed $\hat{\bm{x}}_m$ and the input $\bm{x}$. Meanwhile, the decoder is trained to poorly reconstruct the original input when the CCV does not match the class identity of the input, referred as the non-match condition (${\cdot}_{nm}$). The poorly reconstruction is achieved by minimizing the MSE of the reconstructed $\hat{\bm{x}}_{nm}$ and a randomly sampled signal $\bm{x}_{r}$ from TFIs not belonging to the current class. The loss function is constructed as follows: 

\begin{equation}
\label{loss_re}
\begin{split}
    &\mathcal{L}_{Re} = \mathcal{L}_{Re\_m}+ \mathcal{L}_{Re\_nm} \\
    &= \Vert\hat{\bm{x}}_m-\bm{x}\Vert_2^2+\frac{1}{M-1} \sum_{j=1}^{M-1}\Vert\hat{\bm{x}}_{nm}^j-\bm{x}_r^j\Vert_2^2
\end{split}
\end{equation}

\par When the input signal belongs to a known class, there exists a CCV that matches the latent representations derived from the input, allowing for near-perfect reconstruction of the input image with minimal reconstruction losses. However, for any input of an unknown class not seen during training, its latent representations cannot be adequately matched with any of the available CCV used for decoding. This mismatch renders the reconstruction ineffective for samples from unknown classes and results in significantly higher reconstruction losses. By leveraging this inherent behavior, the magnitude of reconstruction loss can be utilized to discriminate between known and unknown classes. 

\subsubsection{Mutual Information Loss Function}
\par The goal of introducing MI is to learn discriminative and compact latent representations that are optimized for the OSR task. Firstly, maximizing the MI between latent representations and class labels enhances the capture of class-discriminative semantics in the embedded space. As a result, the latent representations become conditioned on class identities rather than low-level input details. Such class-focused representations allow for better matching to CCVs during reconstruction, thereby assisting the role of the class-conditioned reconstruction.

\par Meanwhile, by minimizing the MI between the latent representations and inputs, the proposed CIR disentangles the representations from non-essential variation in the input domain like noise, style differences etc. This helps the model filter out input-specific information irrelevant to class distinctions. This improves generalization by reducing dependence on idiosyncrasies of training examples.



\par The maximizing and minimizing process is achieved by deriving the following objective function \cite{alemi2016deep}.
\begin{equation}\label{eq:mi}
    \mathcal{L}_{MI} = \text{MI} (\bm{Z}, \bm{X}) - \gamma \text{MI} (\bm{Z}, \bm{Y})
\end{equation}
where $\gamma$ is the Lagrange multiplier. $MI(\cdot,\cdot)$ denotes the mutual information with 
\begin{equation}
    \text{MI}(A,B) = \iint p(a,b)\text{log} \left(\frac{p(a,b)}{p(a)p(b)}\right)\text{d}a\text{d}b
\end{equation}


\par However, directly optimizing Equation \ref{eq:mi} is intractable. The MI terms in Equation \ref{eq:mi} are thus estimated separately by optimizing the upper bound of $\text{MI}(\bm{Z}, \bm{X})$ and the lower bound of $\text{MI}(\bm{Z}, \bm{Y})$ as follows:


\begin{equation}\label{eq:mi_im}
    \mathcal{L}_{MI} = \mathcal{L}_{UB}- \gamma \mathcal{L}_{LB}
\end{equation}

\par a) Minimum MI about Input
\par To learn a latent representation that preserves the minimum mutual information about the input data, the CIR minimizes the upper bound motivated by \cite{cheng2020club}, and the target objective is written as:

\begin{equation}
    \mathcal{L}_{UB} =  \mathbb{E}_{p(\bm{Z}, \bm{X})}\left[\text{log} p(x|z)\right]- \mathbb{E}_{p(\bm{Z}) p(\bm{X})}\left[\text{log} p(x|z)\right]
\end{equation}
\par This paper validates that the upper bound can be estimated through the theoretical analysis, and the theorem format is given as follows:

\begin{theorem}
For the random variables $\bm{Z}$ and $\bm{X}$,
\begin{equation}
    \text{MI} (\bm{Z}, \bm{X}) \leq \mathcal{L}_{UB}
\end{equation}
Equality is achieved if and only if $\bm{Z}$ and $\bm{X}$ are independent.
\end{theorem}



\par b) Maximum MI about Class

\par With the goal of obtaining a representation capturing maximum mutual information regarding the class label, the lower bound informed by \cite{belghazi2018mine} is maximized. This yields the following objective:


\begin{equation}
    \mathcal{L}_{LB} =  \mathop{sup}\limits_{F_{\phi}} \mathbb{E}_{p(y,z)}[F_{\phi}(y,z)] -\text{log} \left(\mathbb{E}_{p(y)p(z)}[e^{F_{\phi}(y,z)}] \right)
\end{equation}
where $F_{\phi}: \mathcal{Y} \times \mathcal{Z} \rightarrow \mathcal{R}$ represents several dense layers.

\par This paper validates that the lower bound can be estimated through the theoretical analysis, and the theorem format is given as follows:

\begin{theorem}
For the random variables $\bm{Z}$ and $\bm{Y}$,
\begin{equation}
    \text{MI} (\bm{Z}, \bm{Y}) \geq \mathcal{L}_{LB}
\end{equation}
Equality is achieved if and only if $\bm{Z}$ and $\bm{Y}$ are independent.
\end{theorem}
\par The details of the proof for the two theorem are presented in Appendix \ref{app1}.

\subsection{Training and Inference Procedure}





\par The training procedure of the proposed CIR is summarized in Algorithm \ref{algorithm1}. The algorithm first initializes the parameters and then iterates over the training epochs. In each iteration, the denoising and reconstruction modules are optimized in turn. First, the raw input signal $\bm{x}_{l}$ is processed by the Restormer $\mathcal{N}$ to obtain the de-noised TFI $\bm{x}$. The parameters of denoising module ($\theta_{\mathcal{N}}$) are then updated by minimizing the denoising loss $\mathcal{L}_{De}$. Next, the reconstruction phase begins. The class-conditioned reconstruction procedure is performed with the AE structure and the designed CCVs. The reconstruction losses $\mathcal{L}_{Re\_m}$ and $\mathcal{L}_{Re\_nm}$ are calculated between actual and reconstructed samples for the matched and non-matched classes. Additionally, the mutual information loss  $\mathcal{L}_{MI}$ is computed. These losses drive updates to the parameters of the reconstruction module ($\theta_{\mathcal{F}},\theta_{\mathcal{G}},\theta_{\mathcal{H}_{\alpha}},\theta_{\mathcal{H}_{\beta}}$). 
\par For threshold selection, we set a scenario where 1\% of the known class samples in the training set may be wrongly classified as unknown to get the threshold $\tau$. Given that this study primarily focuses on widening the discrepancy between reconstruction losses of known and unknown classes, a relatively simple decision boundary is sufficient to effectively discriminate between known and unknown samples. To evaluate the impact of threshold selection, we conducted analyses with receiver operating characteristic curves (ROC) obtained by changing $\tau$ in Section \ref{section simu}. 
\begin{algorithm}
\caption{Training Procedure of CIR}\label{algorithm1}
\KwIn{Training datasets $\mathcal{D}_{l}=\{\bm{x}^i_{l},y^i_{l}\}_{i=1}^{N_l}$, epoch number $E$, CCVs $\{l_1,\cdots,l_M\}$}
\KwOut{The well-trained CIR, threshold $\tau$}
\BlankLine
\textit{\# Training Procedure}\\
Initialize the parameters $\theta_{\mathcal{N}},\theta_{\mathcal{F}},\theta_{\mathcal{G}},\theta_{\mathcal{H}_{\alpha}},\theta_{\mathcal{H}_{\beta}}$ of the denoising Restormer $\mathcal{N}$, encoder $\mathcal{F}$, decoder $\mathcal{G}$, fully connected layers $\mathcal{H}_{\alpha},\mathcal{H}_{\beta}$;

\For{$e\leftarrow 1$ \KwTo $E$}{

    Sample $(\bm{x}_{l},y_{l})$ from $\mathcal{D}_{l}$\;
    Sample $\{(\bm{x}^i_{r},y^i_{r})|y^i_{r}\neq y_{l}\}_{i=1}^{M-1}$ from $\mathcal{D}_{l}$ \;
    Obtain denoised TFI by Restormer $\bm{x}=\mathcal{N}(\bm{x}_{l})$\;
    Update $\theta_{\mathcal{N}}$ by descending gradients of losses $\mathcal{L}_{De}$ in Eq. \ref{loss_de}\;
    Obtain latent representations by encoder $\bm{z} = \mathcal{F}(\bm{x})$\;

    \For{$j\leftarrow 1$ \KwTo $M$}{
    Calculate $\bm{\alpha}_j = \mathcal{H}_{\bm{\alpha}}(l_j)$, $\bm{\beta}_j = \mathcal{H}_{\bm{\beta}}(l_j)$\;
    Obtain the class-conditioned latent representations $\bm{z}_j = \bm{\alpha}_j \odot \bm{z} +\bm{\beta}_j$\;
    Obtain class-conditioned reconstruction by decoder $\bm{\hat{x}}_j = \mathcal{G}(\bm{z}_j)$\;
    \eIf{$j==y_l$}{
    Calculate the matched reconstruction loss $ \mathcal{L}_{Re\_m} = \Vert\hat{\bm{x}}_m-\bm{x}\Vert_2^2$\;
    }{
    
    Calculate the non-matched reconstruction loss $\mathcal{L}_{Re\_nm} = \Vert\hat{\bm{x}}_{nm}^j-\bm{x}_r^j\Vert_2^2$\;
    }
    Update $\theta_{\mathcal{F}},\theta_{\mathcal{G}},\theta_{\mathcal{H}_{\alpha}},\theta_{\mathcal{H}_{\beta}}$ by descending gradients of losses $\mathcal{L}_{Re}$ and $\mathcal{L}_{MI}$ in Eq. \ref{loss_re} and Eq. \ref{eq:mi_im}\;
    
    }

  }
  Calculate the matched reconstruction losses for all training samples $\bm{R}=(r_1,r_2,\cdots,r_{N_l})$\;
  Obtain the sorted sequence in ascending order $\bm{R}'$\;
  Threshold  $\tau = \text{min}\{\bm{R}'(i)|i\geq g\}, g = \text{ceil}(0.99N_l)$\;

\end{algorithm}

\par The inference procedure of the proposed CIR is summarized in Algorithm \ref{algorithm2}. The well-trained CIR framework and pre-defined CCVs are utilized to obtain the class-conditioned reconstruction losses set $\bm{R}$ for each test samples. If the minimum class-conditioned reconstruction loss $R_{min}$ exceeds a predefined threshold $\tau$, the sample is identified as belonging to the unknown class. Alternatively, if the minimum loss is below $\tau$, the sample is classified into the class corresponding to that minimal loss.



\begin{algorithm}
\caption{Inference Procedure of CIR}\label{algorithm2}
\KwIn{Testing datasets $\mathcal{D}_{t}=\{\bm{x}^i_{t},y^i_{t}\}_{i=1}^{N_{t}}$,  CCVs $\{l_1,\cdots,l_M\}$, well-trained denoising Restormer $\mathcal{N}$, encoder $\mathcal{F}$, decoder $\mathcal{G}$, fully connected layers $\mathcal{H}_{\alpha},\mathcal{H}_{\beta}$, threshold $\tau$}
\KwOut{Predicted labels $\{\hat{y_i}\}_{i=1}^{N_{t}}$ }

\BlankLine
\textit{\# Testing Procedure}\\
    Obtain denoised TFI by Restormer $\bm{x}=\mathcal{N}(\bm{x}_t)$ \;
    Obtain latent representations by encoder $\bm{z} = \mathcal{F}(\bm{x})$\;
    \For {$j = 1,\cdots,M$}{
    Calculate $\bm{\alpha}_j = \mathcal{H}_{\bm{\alpha}}(l_j)$, $\bm{\beta}_j = \mathcal{H}_{\bm{\beta}}(l_j)$\;
    Obtain the class-conditioned latent representations $\bm{z}_j = \bm{\alpha}_j \odot \bm{z} +\bm{\beta}_j$\;
    Obtain reconstructed TFIs by decoder $\hat{\bm{x}}_j = \mathcal{G}(\bm{z}_j)$\;
    Calculate the class-conditioned reconstruction loss $\bm{R}(j) = \Vert \bm{x} -  \hat{\bm{x}}_j \Vert^2_2$\;
    }

    Obtain the predicted label $\hat{y} = \text{argmin}(\bm{R})$\;
    Obtain $R_{min} = \text{Min}(\bm{R})$\;
    \eIf{${R}_{min}\geq \tau$} 
    {Predict $\bm{x}_t$ as unknown sample with label ${\psi}_{N+1}$\; }
    {Predict $\bm{x}_t$ as known sample with label $\hat{y}$\;}

Performance evaluation\;
\end{algorithm}

\section{Experiments}
\label{section simu}

\subsection{Experiments Design}
\subsubsection{Dataset Description}

\par This paper comprehensively evaluates the performance of the proposed method using both simulated and measured datasets. For the simulated data, 11 modulation types are considered: linear frequency modulation (LFM), Costas code, five polyphase codes (Frank, P1, P2, P3 and P4), and four polytime codes (T1, T2, T3 and T4) \cite{RN107}, with their parameter settings presented in Appendix \ref{app2}. The sampling frequency $f_s$ is set to 50MHz. The carrier frequency $f_c$ is randomly sampled within $(\frac{1}{6},\frac{1}{4})f_s$, and the pulse width is $6\mu s$. The additional white Gaussian noise (AWGN) is considered to simulate real-world conditions, where the SNRs are varied from $-$10 to 10 dB with 2dB increments. 

\par In fact, it is not convincing to evaluate the OSR performance only by testing the methods on the simulated dataset. In this paper, measured signals are obtained by a certain ground-to-air radar to make up a measured signal dataset, which contains signals of five modulation types: Baker7, LFM, NLFM, M15 and Rectangular. The measured environment is at a horn antenna at a height of 8 meters. The pulse repetition period is 400 $\mu s$, and the pulse width varies between 20 and 60 $\mu s$. The intermediate frequency is 2.2 GHz, and the elevation angle of the receiving antenna is approximately $10^{\circ}$. The sampling frequency is 5 GHz, and the maximum bandwidth of the frequency-modulated waveform is 15 MHz. 

\par TFA is conducted on each waveform sample, then the resulted TFIs are downsampled to a smaller size of $80\times 80$. Each modulation type consists of 500 signal samples at each SNR level. These samples are split into training and test datasets in a 4:1 ratio.  

\par For AMOSR problems, the test dataset contains unknown-class samples that are not encountered in the training dataset. Therefore, we need to select specific modulation types as unknown classes to partition the training and test datasets. Depending on the specific experimental objectives, we split the dataset as presented in Table \ref{table6}. For Dataset D1, which focuses on evaluating the robustness of CIR against different SNRs, we selected 7 modulation types as known classes and 4 modulation types as unknown classes. Then, dataset D2 is constructed to validate the performance of CIR across varying levels of openness. Openness is utilized to measure the ratio of known classes and unknown classes, defined as 
\begin{equation}
    \text{Openness} = 1-\sqrt{\frac{N}{N+N_u}}
\end{equation}
As the number of unknown modulation types increases, the degree of openness increases. Lastly, dataset D3 constructs five subsets from the measured signals to evaluate the AMOSR performance, including three subsets with single unknown class and two subsets with a pair of unknown classes.

\begin{table*}[htbp]
  \centering
  \caption{Dataset Configuration}

    \begin{tabular}{llll}
    \toprule
    \multicolumn{1}{p{5em}}{Datasets} & \multicolumn{1}{p{8em}}{SNR} & \multicolumn{1}{p{18em}}{Training Datasets} & \multicolumn{1}{p{14em}}{Testing Datasets (UNK)} \\
        \midrule
    \multicolumn{1}{l}{\multirow{2}[2]{*}{\thead{D1\\(Simulated)}}} & \multicolumn{1}{l}{\multirow{2}[2]{*}{\thead{[-10,-8-6,-4,-2,\\2,4,6,8,10]dB}}} & \multicolumn{1}{l}{\multirow{2}[2]{*}{LFM\textbackslash{}P1$-$P4\textbackslash{}T1\textbackslash{}T3}} & \multicolumn{1}{l}{\multirow{2}[2]{*}{Frank\textbackslash{}Costas\textbackslash{}T2\textbackslash{}T4}} \\
          &       &       &  \\
    \midrule
    \multicolumn{1}{l}{\multirow{10}[2]{*}{\thead{D2\\(Simulated)}}} & \multicolumn{1}{l}{\multirow{10}[2]{*}{[-8,-4,0,4]dB}} & \multicolumn{1}{p{13.75em}}{LFM\textbackslash{}Costas\textbackslash{}Frank\textbackslash{}P2$-$P4\textbackslash{}T1$-$T4} & \multicolumn{1}{p{13.565em}}{P1} \\
          &       & \multicolumn{1}{p{13.75em}}{LFM\textbackslash{}Costas\textbackslash{}Frank\textbackslash{}P3\textbackslash{}P4\textbackslash{}T1$-$T4} & \multicolumn{1}{p{13.565em}}{P1\textbackslash{}P2} \\
          &       & \multicolumn{1}{p{13.75em}}{LFM\textbackslash{}Costas\textbackslash{}Frank\textbackslash{}P3\textbackslash{}P4\textbackslash{}T2$-$T4} & \multicolumn{1}{p{13.565em}}{P1\textbackslash{}P2\textbackslash{}T1} \\
          &       & \multicolumn{1}{p{13.75em}}{LFM\textbackslash{}Costas\textbackslash{}Frank\textbackslash{}P3\textbackslash{}P4\textbackslash{}T2\textbackslash{}T4} & \multicolumn{1}{p{13.565em}}{P1\textbackslash{}P2\textbackslash{}T1\textbackslash{}T3} \\
          &       & \multicolumn{1}{p{13.75em}}{LFM\textbackslash{} Frank\textbackslash{}P3\textbackslash{}P4\textbackslash{}T2\textbackslash{}T4} & \multicolumn{1}{p{13.565em}}{Costas\textbackslash{}P1\textbackslash{}P2\textbackslash{}T1\textbackslash{}T3} \\
          &       & \multicolumn{1}{p{13.75em}}{LFM\textbackslash{}P3\textbackslash{}P4\textbackslash{}T2\textbackslash{}T4} & \multicolumn{1}{p{13.565em}}{Costas\textbackslash{}Frank\textbackslash{}P1\textbackslash{}P2\textbackslash{}T1\textbackslash{}T3} \\
          &       & \multicolumn{1}{p{13.75em}}{LFM\textbackslash{}P4\textbackslash{}T2\textbackslash{}T4} & \multicolumn{1}{p{13.565em}}{Costas\textbackslash{}Frank\textbackslash{}P1$-$P3\textbackslash{}T1\textbackslash{}T3} \\
          &       & \multicolumn{1}{p{13.75em}}{LFM\textbackslash{}T2\textbackslash{}T4} & \multicolumn{1}{p{13.565em}}{Costas\textbackslash{}Frank\textbackslash{}P1$-$P4\textbackslash{}T1\textbackslash{}T3} \\
          &       & \multicolumn{1}{p{13.75em}}{LFM\textbackslash{}T4} & \multicolumn{1}{p{13.565em}}{Costas\textbackslash{}Frank\textbackslash{}P1$-$P4\textbackslash{}T1$-$T3} \\
          &       & \multicolumn{1}{p{13.75em}}{LFM} & \multicolumn{1}{p{13.565em}}{Costas\textbackslash{}Frank\textbackslash{}P1$-$P4\textbackslash{}T1$-$T4} \\
    \midrule
    \multicolumn{1}{l}{\multirow{5}[2]{*}{\thead{D3\\(Measured)}}} & \multicolumn{1}{l}{\multirow{5}[2]{*}{33dB}} & \multicolumn{1}{p{13.75em}}{NLFM\textbackslash{}Baker7\textbackslash{}M15\textbackslash{}Rectangular} & \multicolumn{1}{p{13.565em}}{LFM} \\
          &       & \multicolumn{1}{p{13.75em}}{LFM\textbackslash{}Baker7\textbackslash{}M15\textbackslash{}Rectangular} & \multicolumn{1}{p{13.565em}}{NLFM} \\
          &       & \multicolumn{1}{p{13.75em}}{LFM\textbackslash{}NLFM\textbackslash{}M15\textbackslash{}Rectangular} & \multicolumn{1}{p{13.565em}}{Baker7} \\
          &       & \multicolumn{1}{p{13.75em}}{LFM\textbackslash{}NLFM\textbackslash{}Rectangular} & \multicolumn{1}{p{13.565em}}{M15\textbackslash{}Baker7} \\
          &       & \multicolumn{1}{p{13.75em}}{M15\textbackslash{}Baker7\textbackslash{}Rectangular} & \multicolumn{1}{p{13.565em}}{LFM\textbackslash{}NLFM} \\

    \bottomrule
    \end{tabular}%

  \label{table6}%
\end{table*}%

\subsubsection{Baseline Method}

\par In this paper, we have chosen three representative OSR method in CV field and three sota OSR approaches in AMR field as baseline methods. The selected approaches are described below:

\begin{itemize}
    \item CGDL: Conditional Gaussian Distribution Learning is a method that utilizes VAE to learn class conditional posterior distributions for unknown detection \cite{RN104}.
 
    \item Openmax: OpenMax involves training a classifier to recognize known categories, computing the open-set probability for each class, and subsequently determining a suitable threshold to identify unknown categories.\cite{RN105}.

    
    \item CROSR: Classification-Reconstruction learning for Open-Set Recognition trains networks for joint classification and reconstruction and utilizes latent representations for unknown detection \cite{RN103}.

    \item SR2CNN: Signal Recognition and reconstruction Convolutional Neural Networks is a zero-shot learning framework, which learns the representation of signal semantic feature space for signal recognition \cite{RN504}.

    \item ICS-T: An AMOSR method combined with improved Transformer and the modified Intra-Class Splitting approaches \cite{RN113}.

    \item MT-CGAN: A novel structured extension of the Counterfactual GAN, which utilizes Multi-task Learning strategy to take advantage of the modulation-domain information from the captured signal \cite{RN209}.

\end{itemize}

\par To ensure a fair comparison, we confirm that the parameters of the above baseline methods are fine-tuned at their best state for the experimental dataset.

\subsubsection{Evaluation Metrics}
\par 
\par a) Metrics for KKC
\par The recognition accuracy of known samples (AKS) can be defined as:
\begin{equation}
    \text{AKS} = \frac{\sum_{i=1}^{N}(\text{TP}_i+\text{TN}_i)}{\sum_{i=1}^{N}(\text{TP}_i+\text{TN}_i+\text{FP}_i+\text{FN}_i)}
\end{equation}
where $N$ represents the number of known classes. TP$_i$ and FP$_i$ represent the number of correctly and incorrectly classified positive samples of the $i$-th class, respectively. TN$_i$ and FN$_i$ represent the number of correctly and incorrectly classified negative samples of the $i$-th class, respectively.
\par b) Metrics for UCI
\par The rejection accuracy of unknown samples (AUS) can be defined as:
\begin{equation}
    \text{AUS} = \frac{\text{TU}}{\text{TU}+\text{FU}}
\end{equation}
where TU represents the number of correctly identified unknown samples, while FU represents the number of incorrectly identified unknown samples.


\begin{equation}
    \text{TPR} = \frac{\text{TU}}{\text{TU} + \text{FK}}
\end{equation}
\par False Positive Rate (FPR) is a metric defined as:

\begin{equation}
    \text{FPR} = \frac{\text{FU}}{\text{FU} + \text{TK}}
\end{equation}
where TK represents the number of correctly identified known samples, while FK represents the number of incorrectly identified known samples.

\par c) Metrics for overall performance
\par The normalized accuracy (NA) weights the AKS and AUS, denoted as:
\begin{equation}
    \text{NA} = \lambda \text{AKS}+(1-\lambda)\text{AUS}
\end{equation}
where $\lambda$ represents the proportion of known class samples in the total number of samples.

\par The macro F-measure is defined as a harmonic mean of precision P and recall R, denoted as:

\begin{equation}
    \text{F} = 2\times \frac{\text{P}\times \text{R}}{\text{P}+\text{R}}
\end{equation}

\begin{equation}
    \text{P} = \frac{1}{N}\sum_{i=1}^N\frac{\text{TP}_i}{\text{TP}_i+\text{FP}_i+\text{FU}}
\end{equation}

\begin{equation}
    \text{R} = \frac{1}{N}\sum_{i=1}^N\frac{\text{TP}_i}{\text{TP}_i+\text{FN}_i+\text{FU}}
\end{equation}


\subsection{Performance and Analysis}

\subsubsection{Ablation Study}
\label{section abla}

\par The proposed method introduces Restormer in the denoising module, aiming to improve performance at low SNRs. Additionally, CCVs and MI loss function are elaborately designed to enhance the discrepancy in reconstruction losses between known and unknown classes, thereby achieving better UCI performance. Ablation studies are conducted to assess the contributions of the aforementioned design elements. The following configurations are examined: 1) Configuration without denoising module (n-Denoise); 2) Configuration without CCVs (n-CCV); 3) Configuration without MI loss function (n-MI).

\par
\begin{figure}[!t]
      \centering
      \subfloat[The KCC performance]{\includegraphics[width=1.72in]{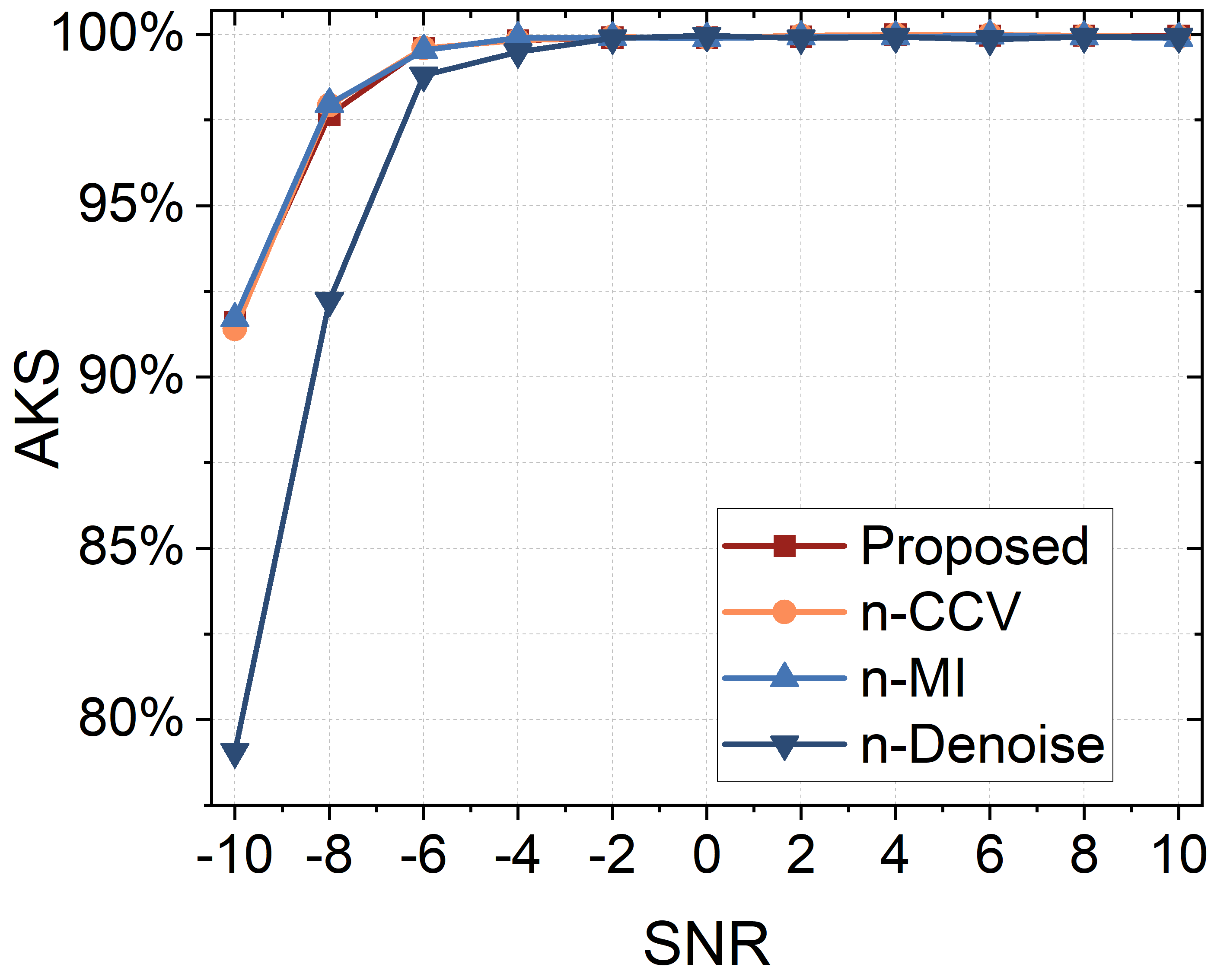}}
      \label{figb2a}
      \subfloat[The UCI performance]{\includegraphics[width=1.72in]{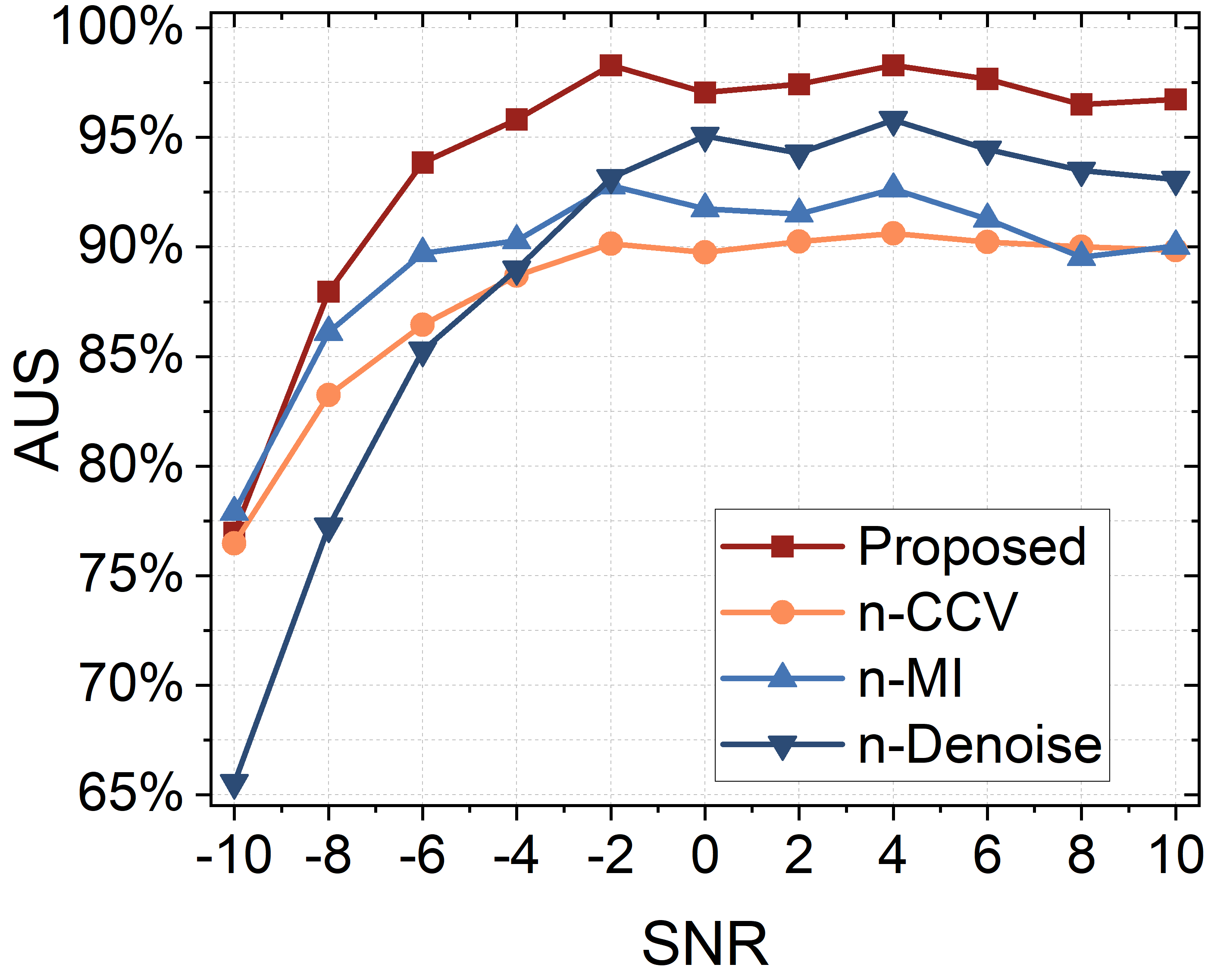}}
      \label{figb2b}
      \captionsetup{justification=justified}
      \caption{Ablation performance.}
      \label{figb2}
  \end{figure}

 The KCC and UCI performance with different configurations at different SNRs are presented in Fig. \ref{figb2}. For the KCC task, when the SNR is above $-$4dB, all configurations achieve an AKS of over 99\%. However, as the SNR decreases, it becomes evident that n-Denoise experiences a noticeable decline in performance. At -10dB, the AKS of the n-Denoise is 12.54 percentage points (pp) lower than the AKS of others. By reducing the noise level, the denoising module improves the quality of the signal, thereby enhancing the KCC performance effectively. For the UCI task, the proposed CIR consistently exhibits superior performance across a range of SNRs. Interestingly, when the SNR drops below -6dB, the AUS of n-Denoise is significantly lower than that of n-CCV and n-MI. However, as the SNR increases, the AUS of n-Denoise surpasses that of n-CCV and n-MI. These results underscore that denoising module contributes enormously to UCI task at low SNRs but the CCVs and MI loss function play the crucial role in improving UCI performance at higher SNRs. 

\par
\begin{figure}[!t]
      \centering
      \subfloat[SNR=8dB]{\includegraphics[width=1.72in]{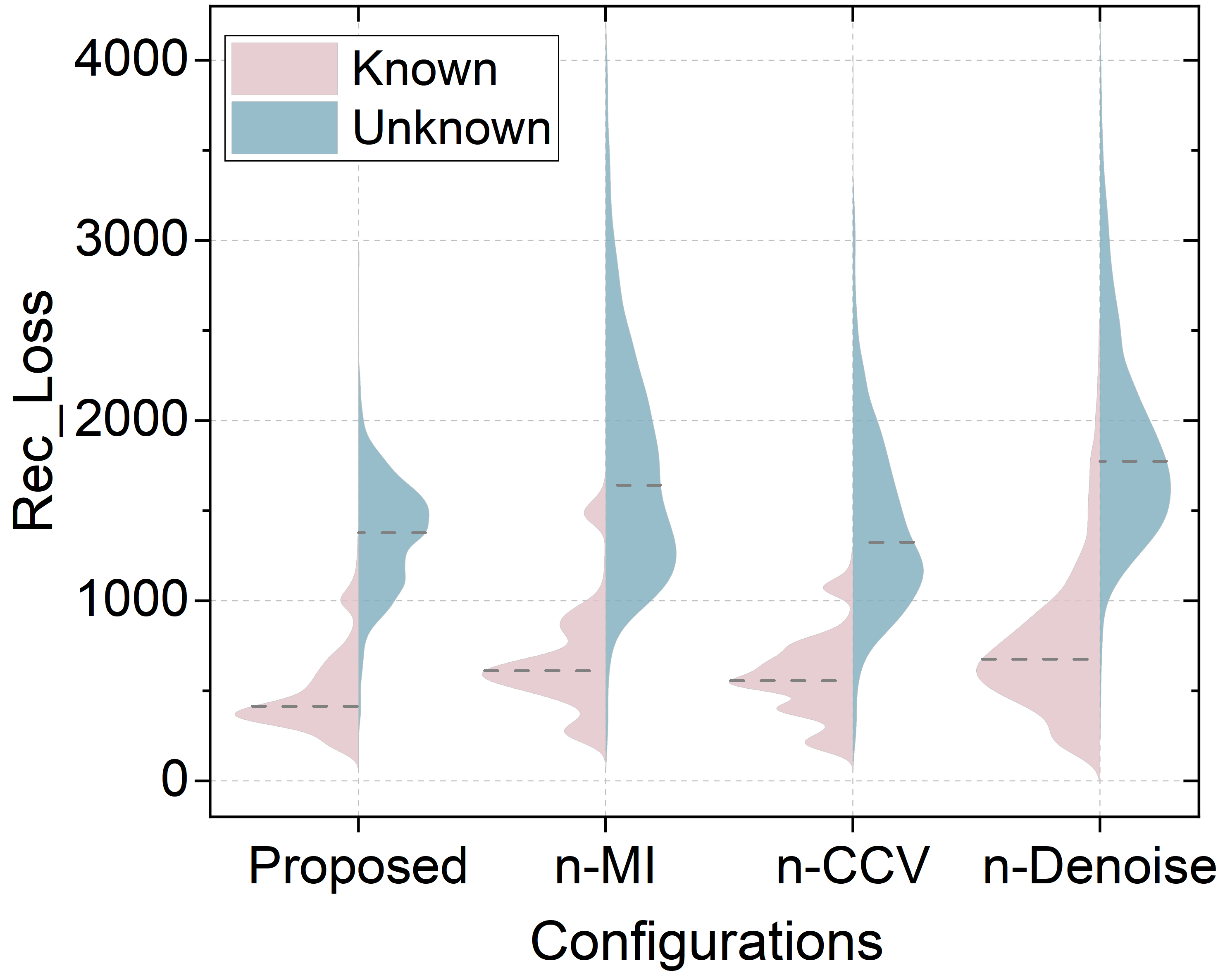}}
      \label{figb2a}
      \subfloat[SNR=4dB]{\includegraphics[width=1.72in]{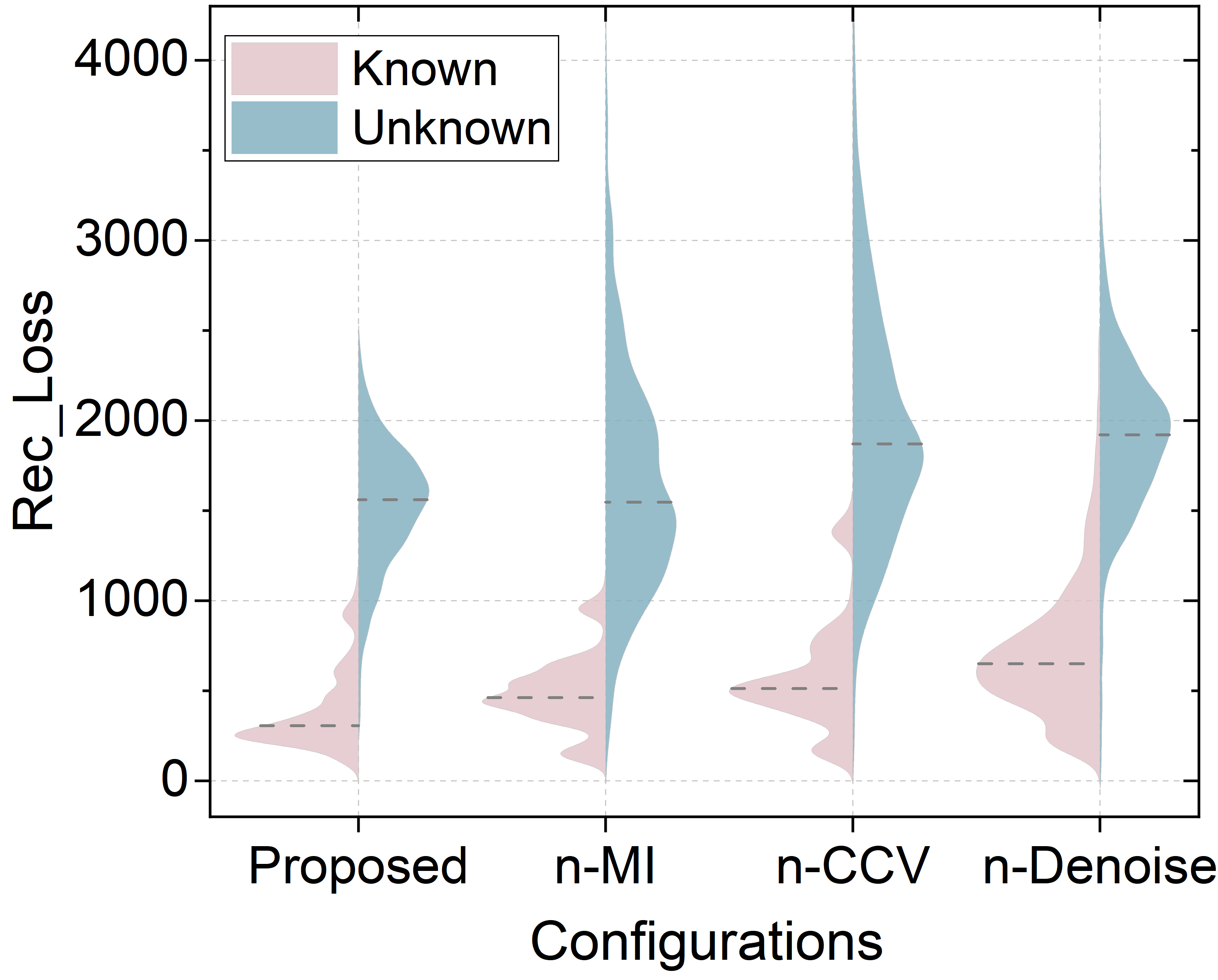}}
      \label{figb2b}
      \subfloat[SNR=-4dB]{\includegraphics[width=1.72in]{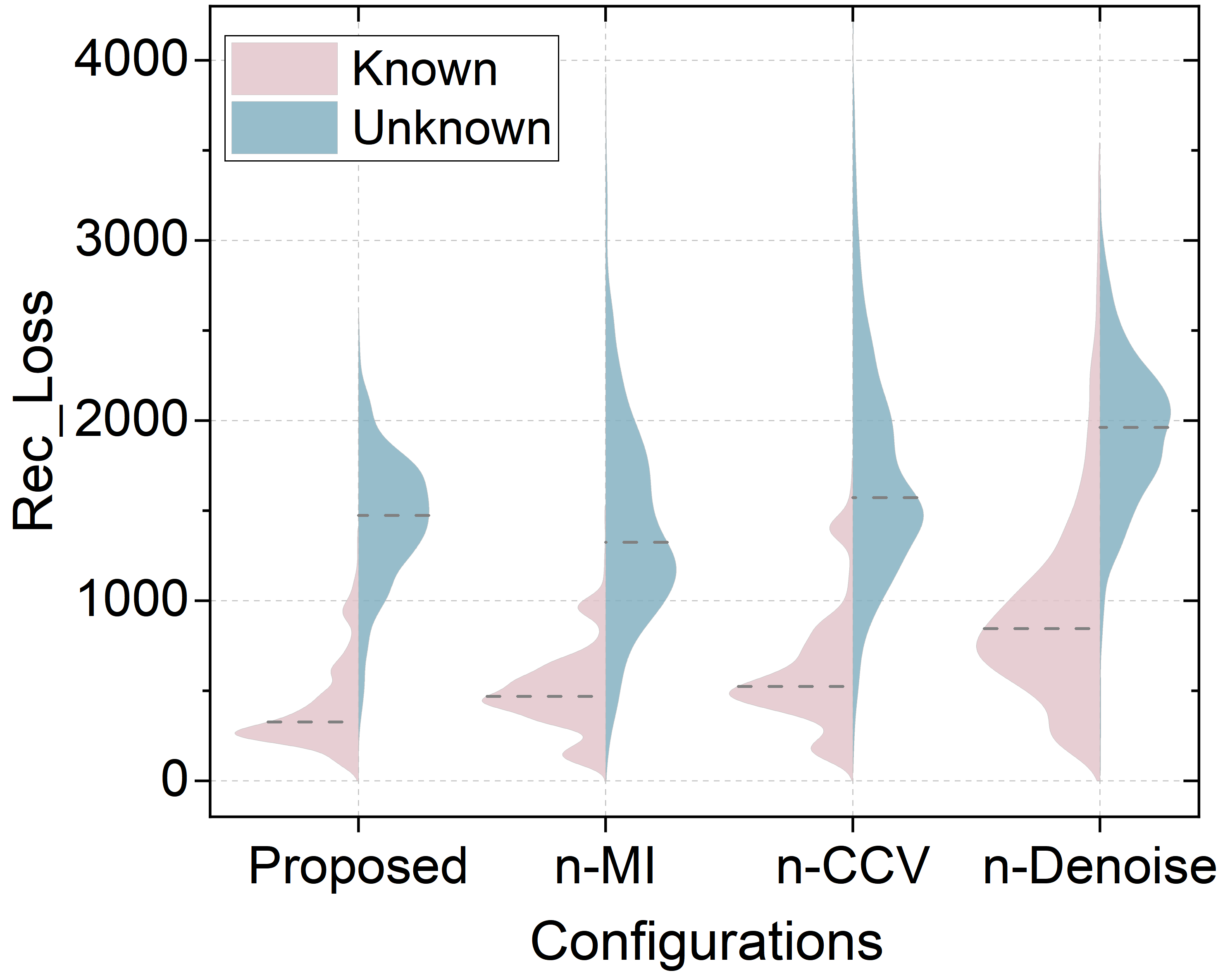}}
      \label{figb2b}
      \subfloat[SNR=-8dB]{\includegraphics[width=1.72in]{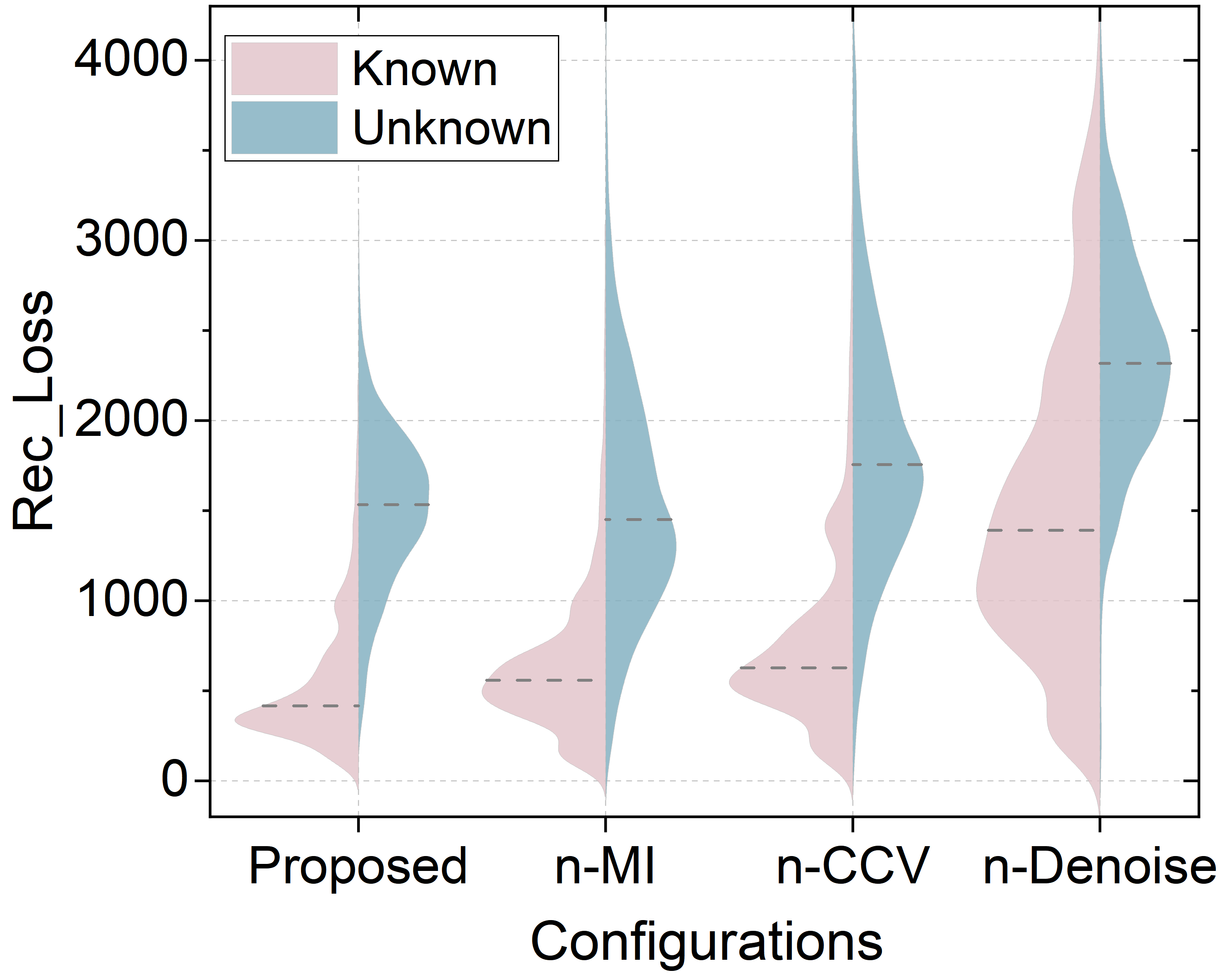}}
      \label{figb2b}
      \captionsetup{justification=justified}
      \caption{The distribution of reconstruction losses for known and unknown classes at different SNRs.}
      \label{figb2_2}
  \end{figure}

 To further investigate, we analyze the distributions of reconstruction losses at different SNRs in Fig. \ref{figb2_2}. It can be observed that when the SNR $\geq$ -4dB, the proposed method exhibits more prominent differences between the distributions of reconstruction losses for known and unknown classes, with the distributions being more concentrated. In contrast, for the n-MI and n-CVV, the peak of the distribution for unknown class reconstruction losses is closer to that of the known class. As for n-Denoise, the distributions of reconstruction losses for known and unknown classes are more dispersed, resulting in greater confusion between the two. When the SNR is -8dB, n-Denoise leads to increased overlap between known and unknown distributions due to greater spread, highlighting the importance of denoising under low SNR conditions.

\subsubsection{Performance Against SNR}


  \begin{figure}[!t]
      \centering
      \subfloat[The KCC performance]{\includegraphics[width=1.72in]{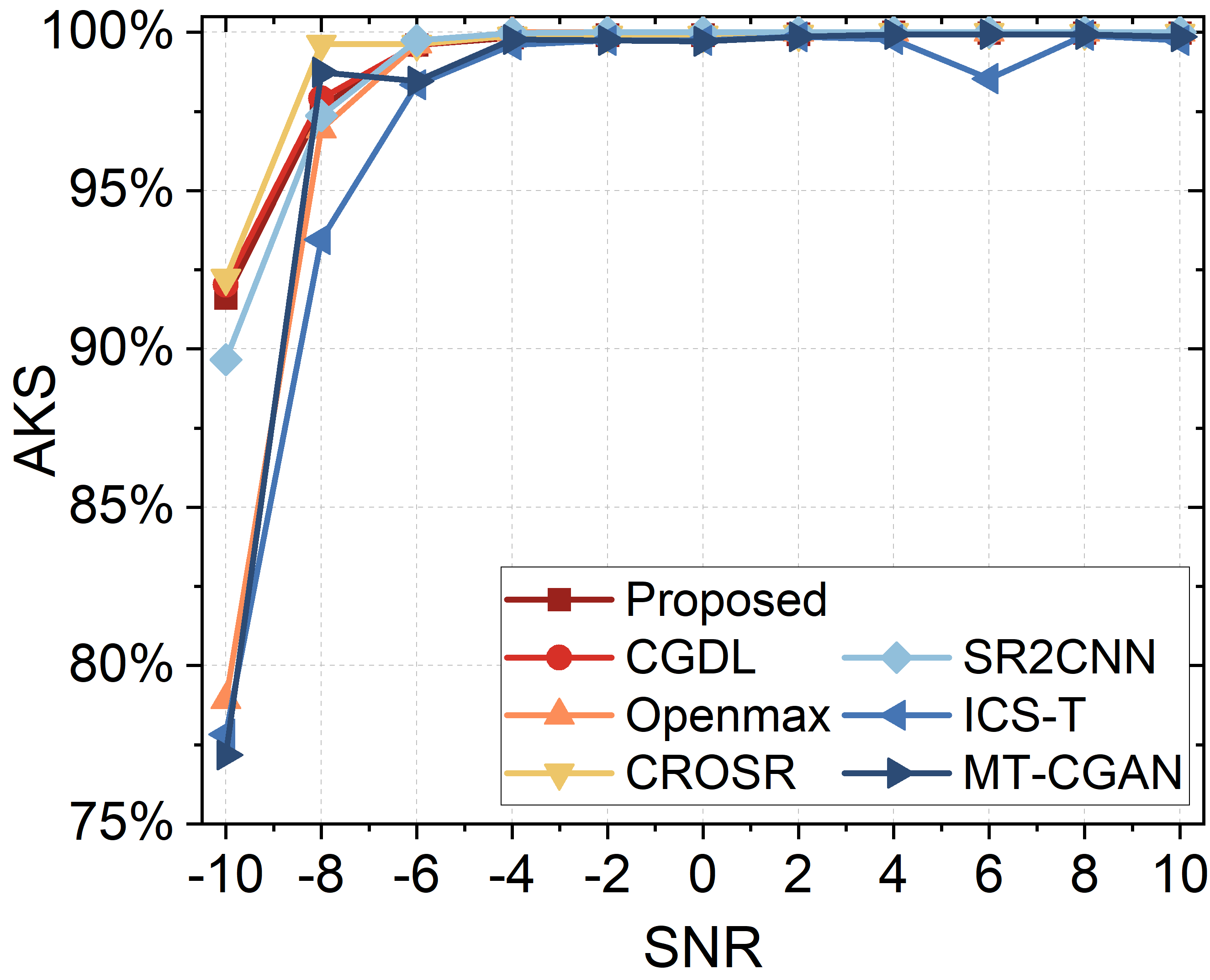}}
      \label{figb3a}
      \subfloat[The UCI performance]{\includegraphics[width=1.72in]{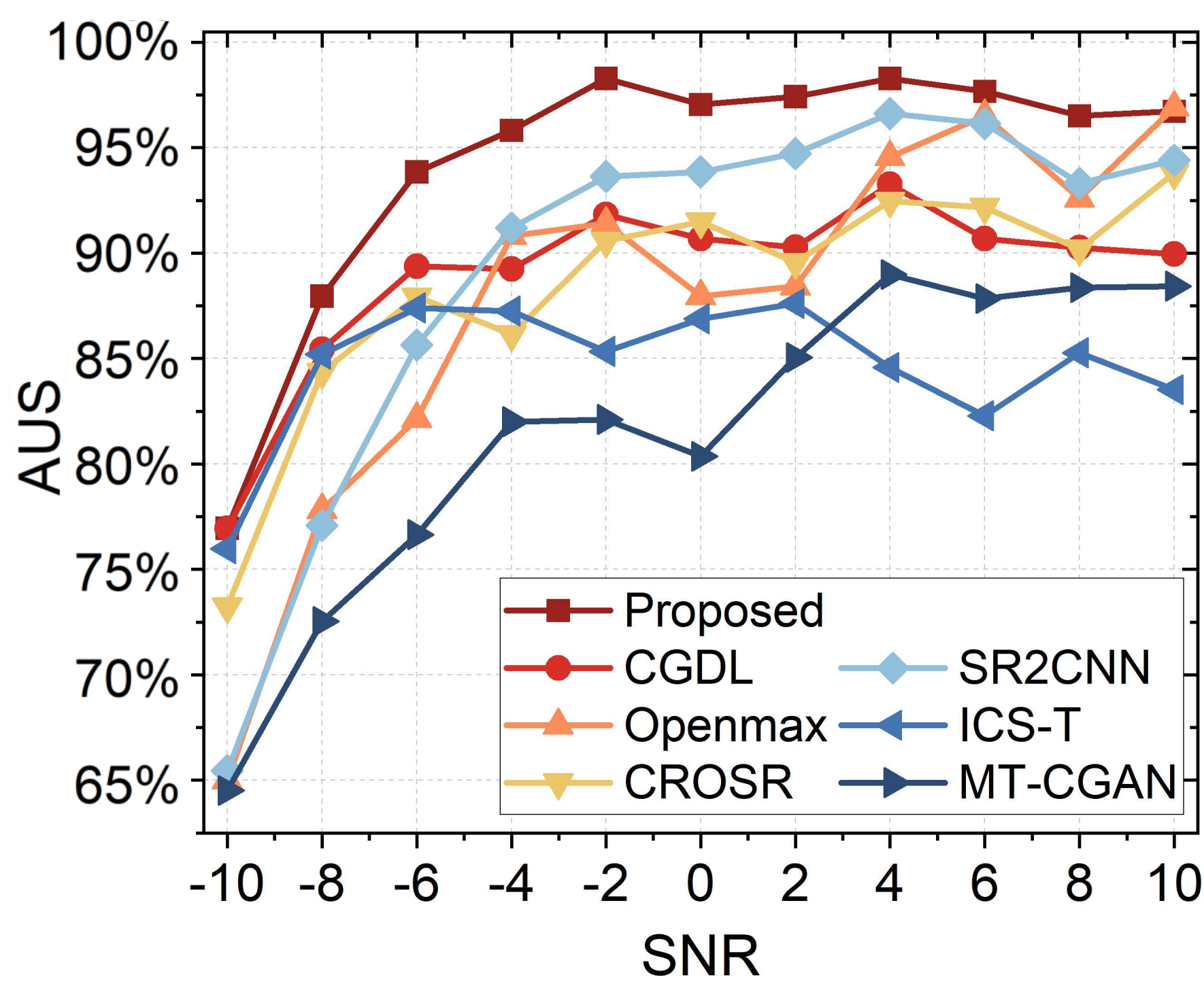}}
      \label{figb3b}
      \captionsetup{justification=justified}
      \caption{The performance of different approaches against SNRs.}
      \label{figb3}
  \end{figure}

    \begin{figure}[!t]
      \centering
      \subfloat[SNR=4dB]{\includegraphics[width=1.72in]{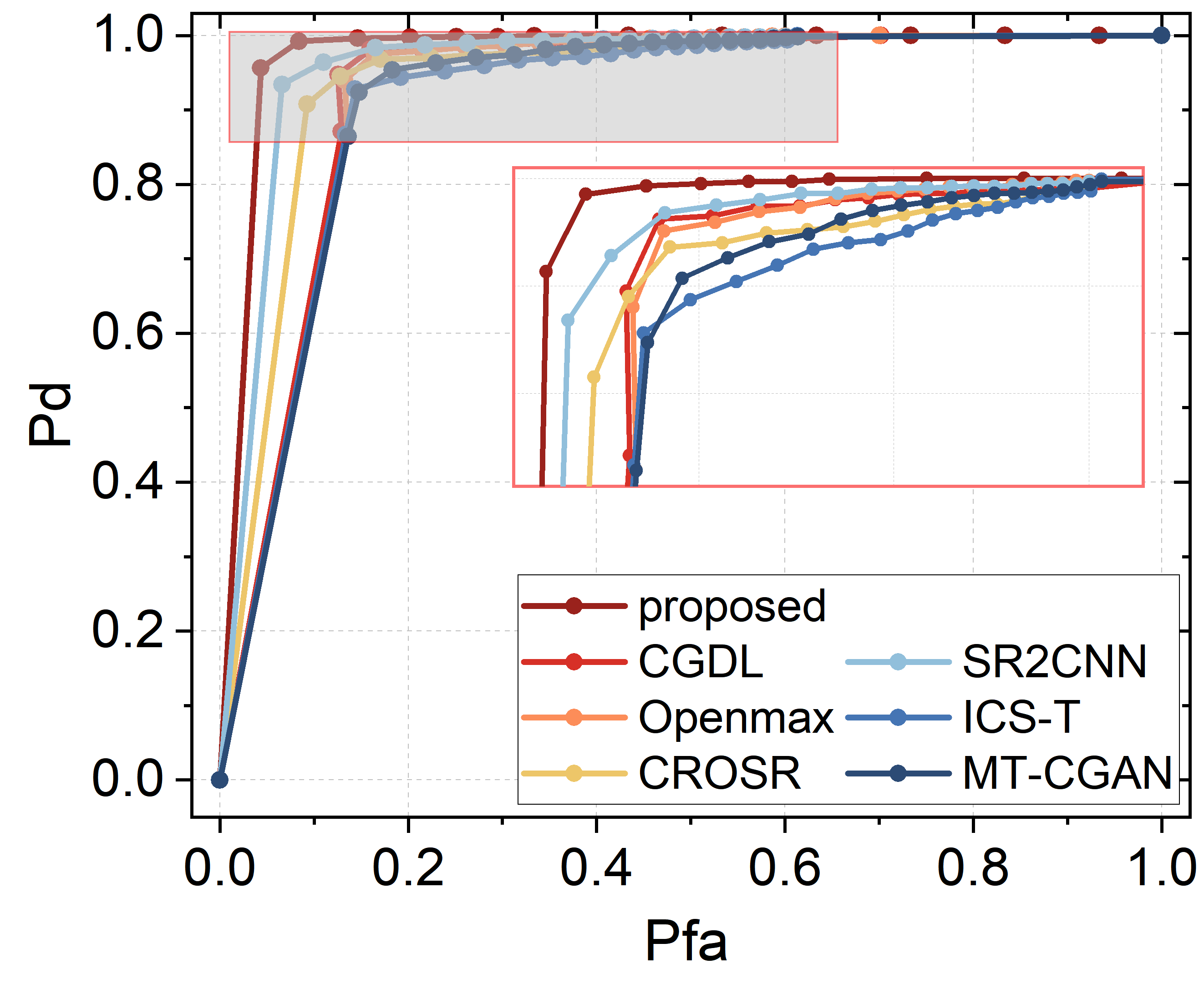}}
      \label{figb2a}
      \subfloat[SNR=0dB]{\includegraphics[width=1.72in]{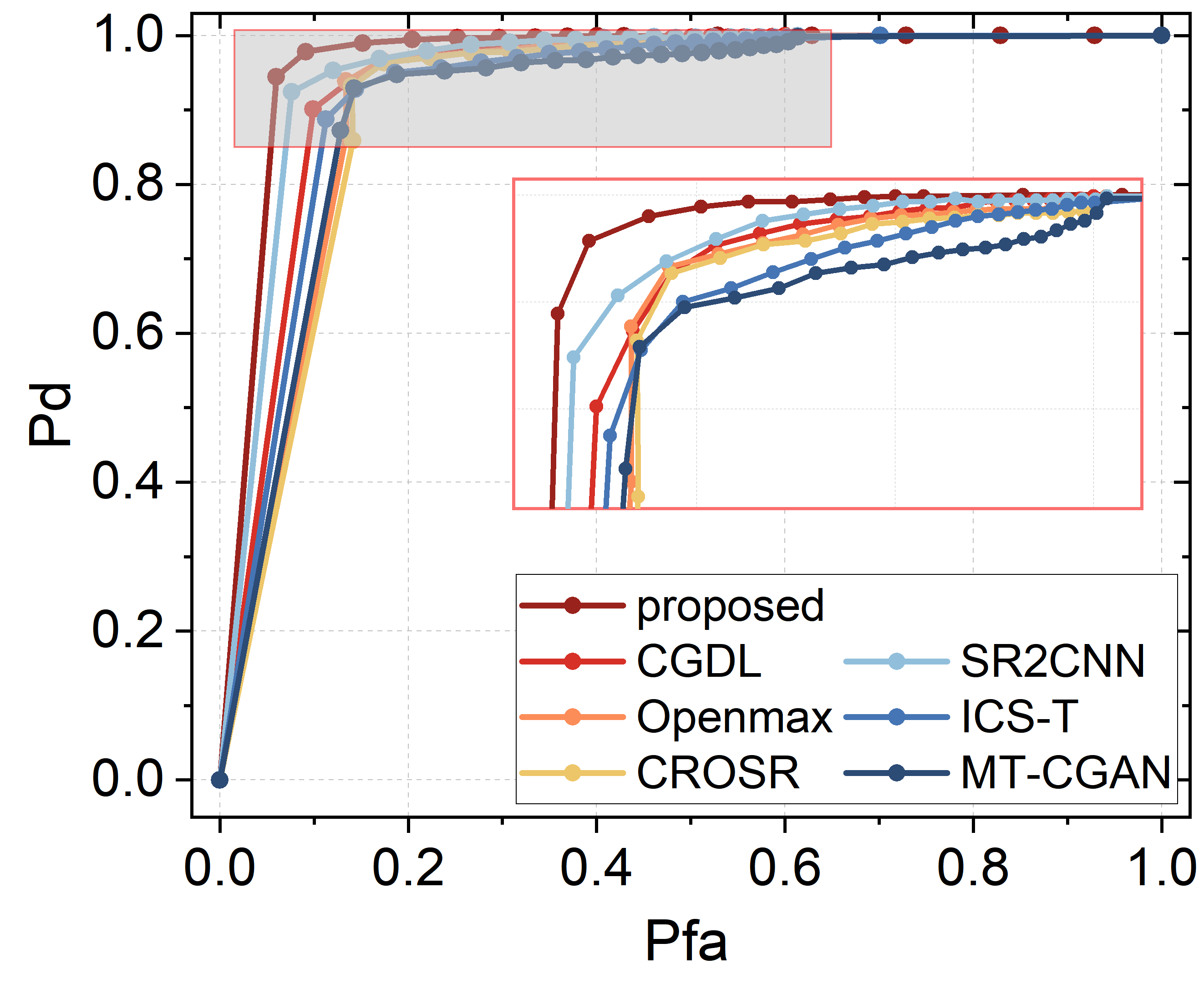}}
      \label{figb2b}
      \subfloat[SNR=-4dB]{\includegraphics[width=1.72in]{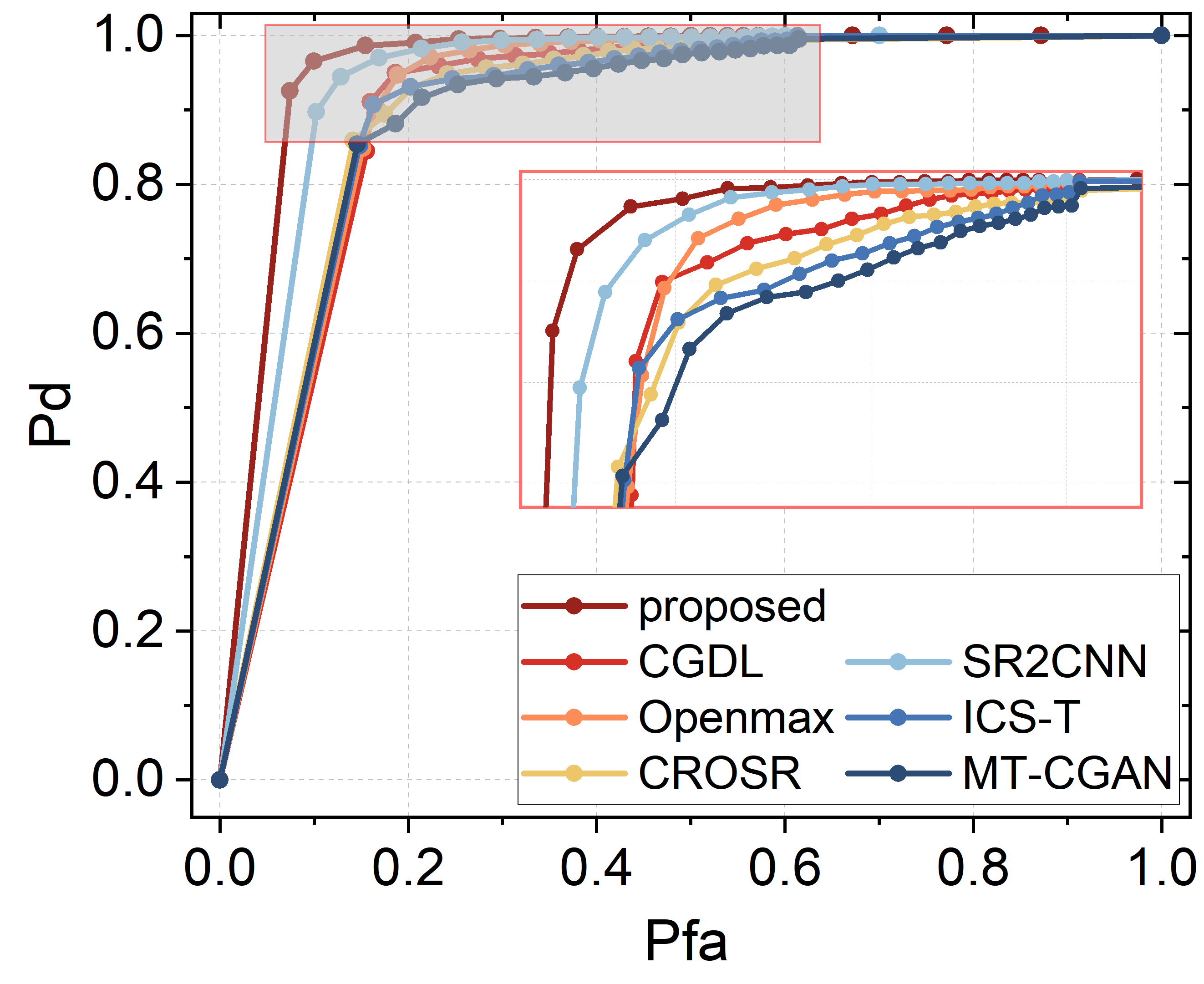}}
      \label{figb2b}
      \subfloat[SNR=-8dB]{\includegraphics[width=1.72in]{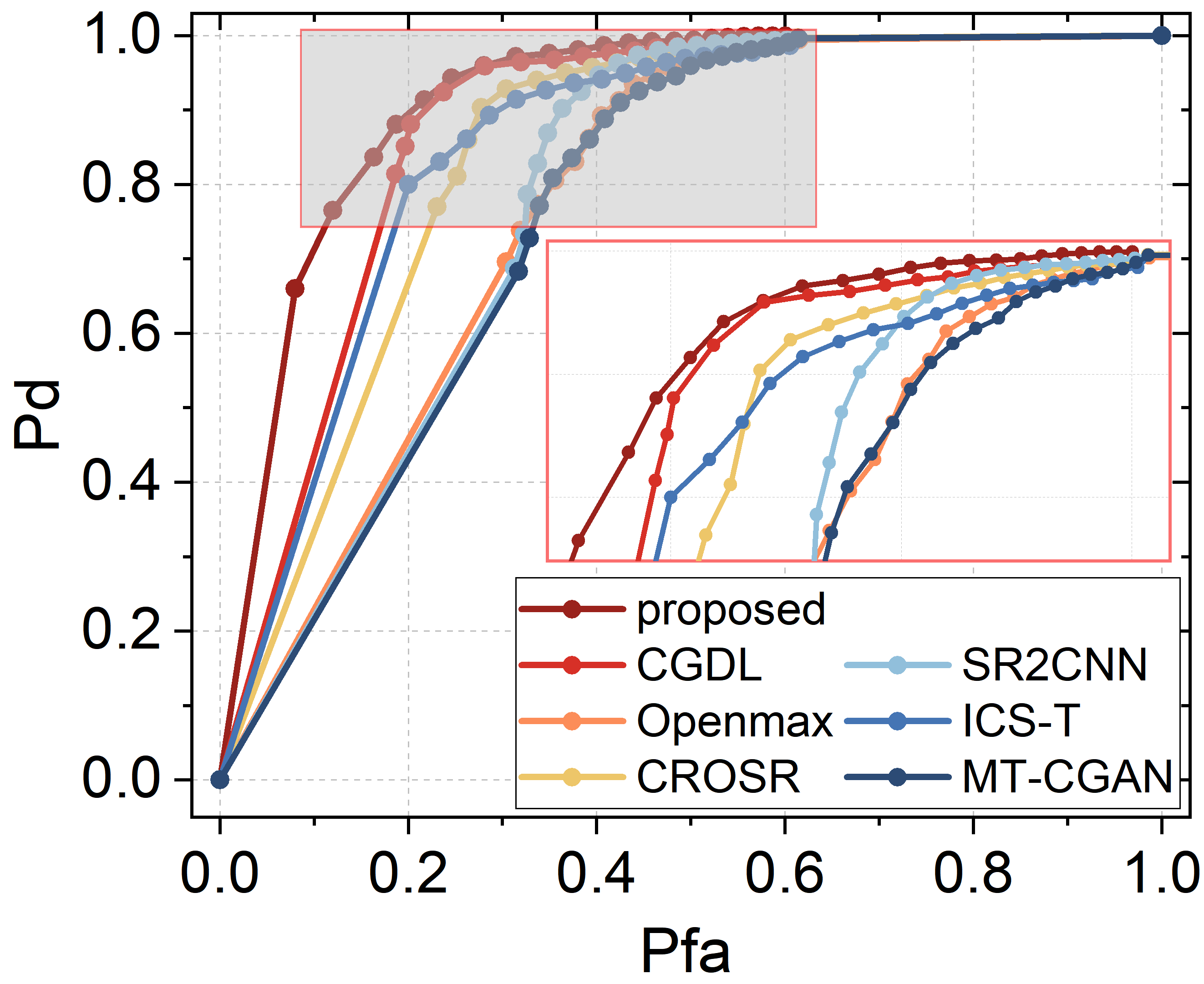}}
      \label{figb2b}
      \captionsetup{justification=justified}
      \caption{The ROC curves of different approaches at different SNRs.}
      \label{figb2_3}
  \end{figure}

\par The Fig. \ref{figb3} illustrates the performance of the proposed method and baseline methods at different SNRs. For the KCC task, all methods achieve high AKS above 99\% when the SNR is higher than -4dB. As the SNR decreases, Openmax, ICS-T, MT-CGAN exhibit a significant drop in AKS compared to the other methods. It can be observed that the proposed method, CGDL and CROSR are all AE-based method, showing better KCC performance at low SNRs. This is because the AE compresses the input TFI into a low-dimensional representation and attempts to reconstruct the original input from this representation. This process forces the AE to learn the key features in the data, thereby enhancing the representation capability of the classifier.

\par For the UCI task, the proposed method demonstrates the best performance across a range of SNRs. In particular, within the SNR range of -8dB to 2dB, the proposed method outperforms the best performing baseline on average by over 5pp. This suggests the stronger robustness of proposed approach against noise perturbations compared to other methods. Interestingly, the performance of the ICS-T method does not exhibit a clear trend with changing SNR. This can be attributed to the fact that ICS relies on the relative density distribution of known class samples. Since noise does not affect the relationships between samples, the boundary determined by ICS remains stable even as the input SNR changes. 

\par To further compare the performance of the proposed method against baselines, Fig. \ref{figb2_3} shows the Receiver Operating Characteristic (ROC) curves obtained at different SNR levels. Across all SNRs considered, our method achieves the best classification accuracy for any false positive rate, as seen from the ROC curves located closest to the top-left corner in each sub-figure. Furthermore, as the false positive rate decreases, the advantage of our method over the baselines in terms of true positive rate (i.e. detection probability of unknowns) becomes more prominent. This demonstrates the proposed approach attains significantly higher detection power for unknown classes even at very low false alarm rates.

\subsubsection{Performance Against Openness}

  \begin{figure}[!t]
      \centering
      \subfloat[SNR=4dB]{\includegraphics[width=1.72in]{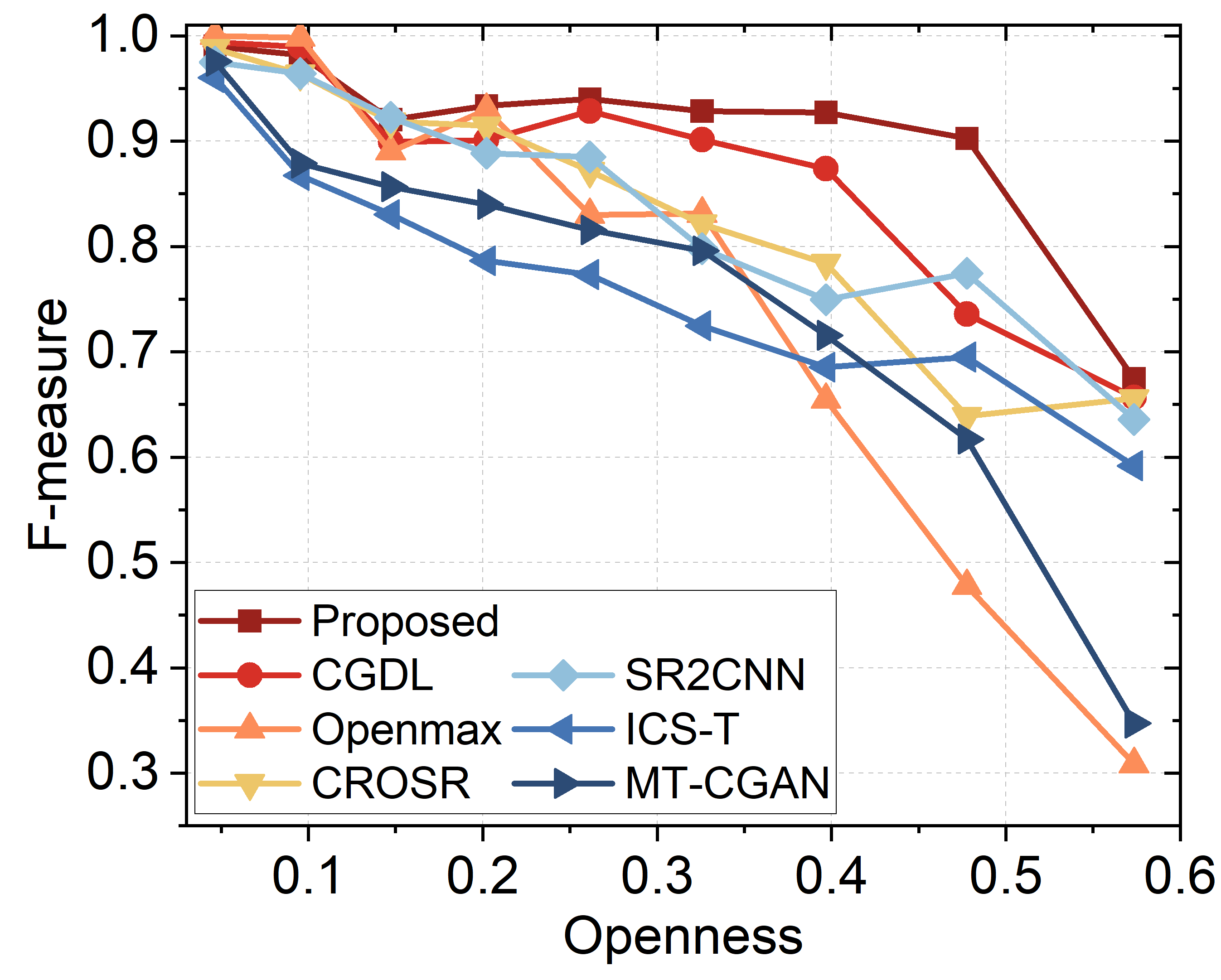}}
      \label{figb4a}
      \subfloat[SNR=0dB]{\includegraphics[width=1.72in]{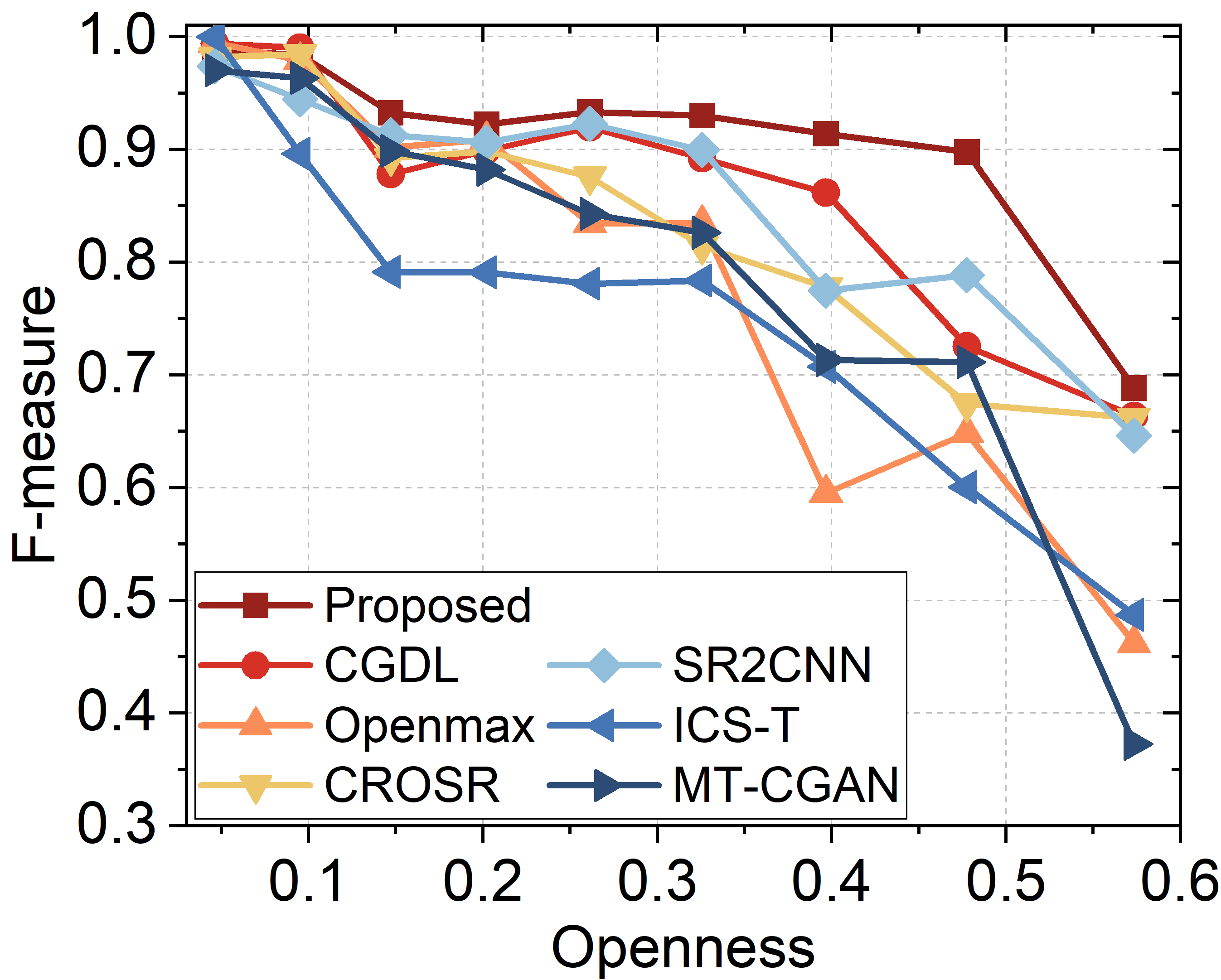}}
      \label{figb4b}
      \subfloat[SNR=-4dB]{\includegraphics[width=1.72in]{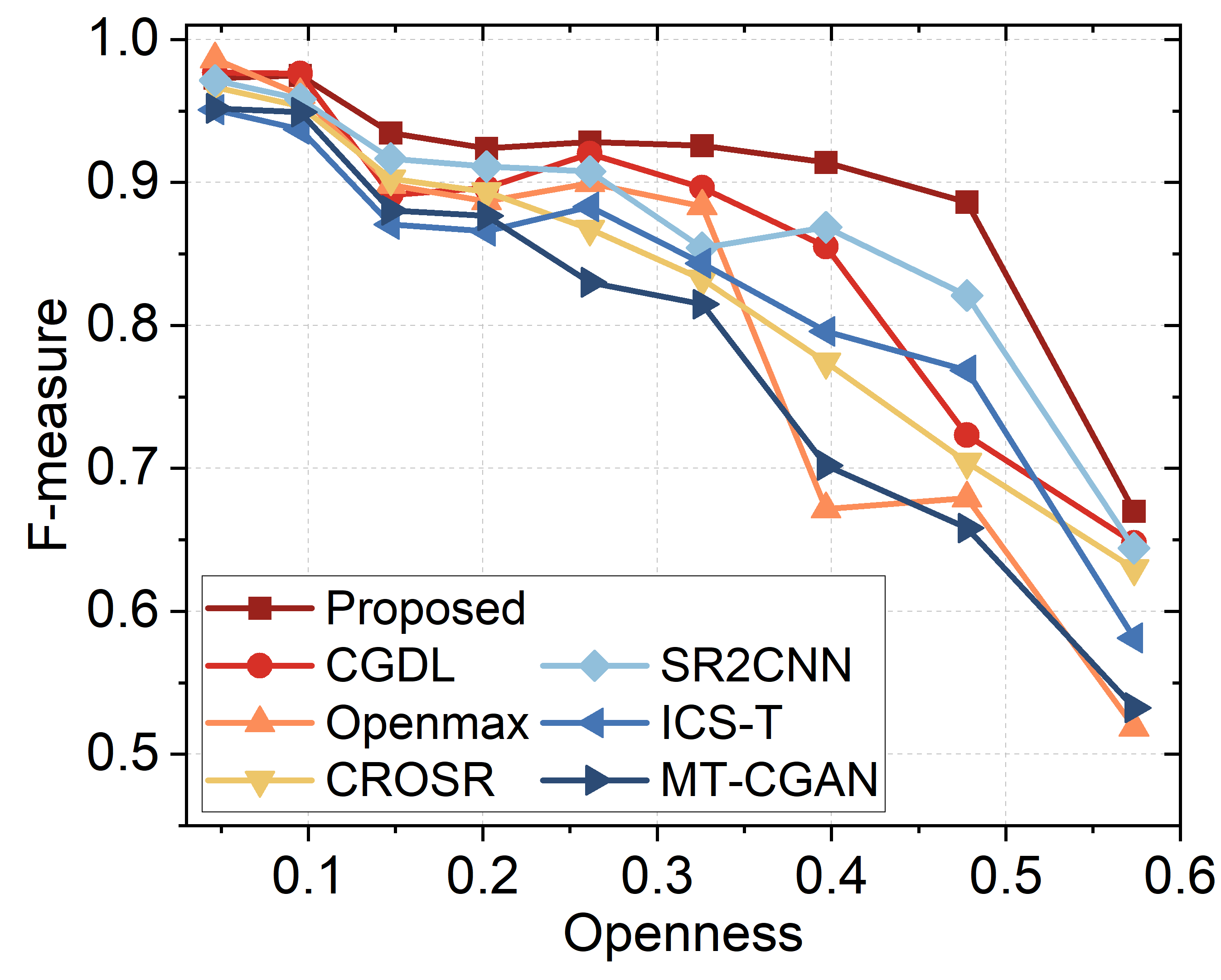}}
      \label{figb4c}
      \subfloat[SNR=-8dB]{\includegraphics[width=1.72in]{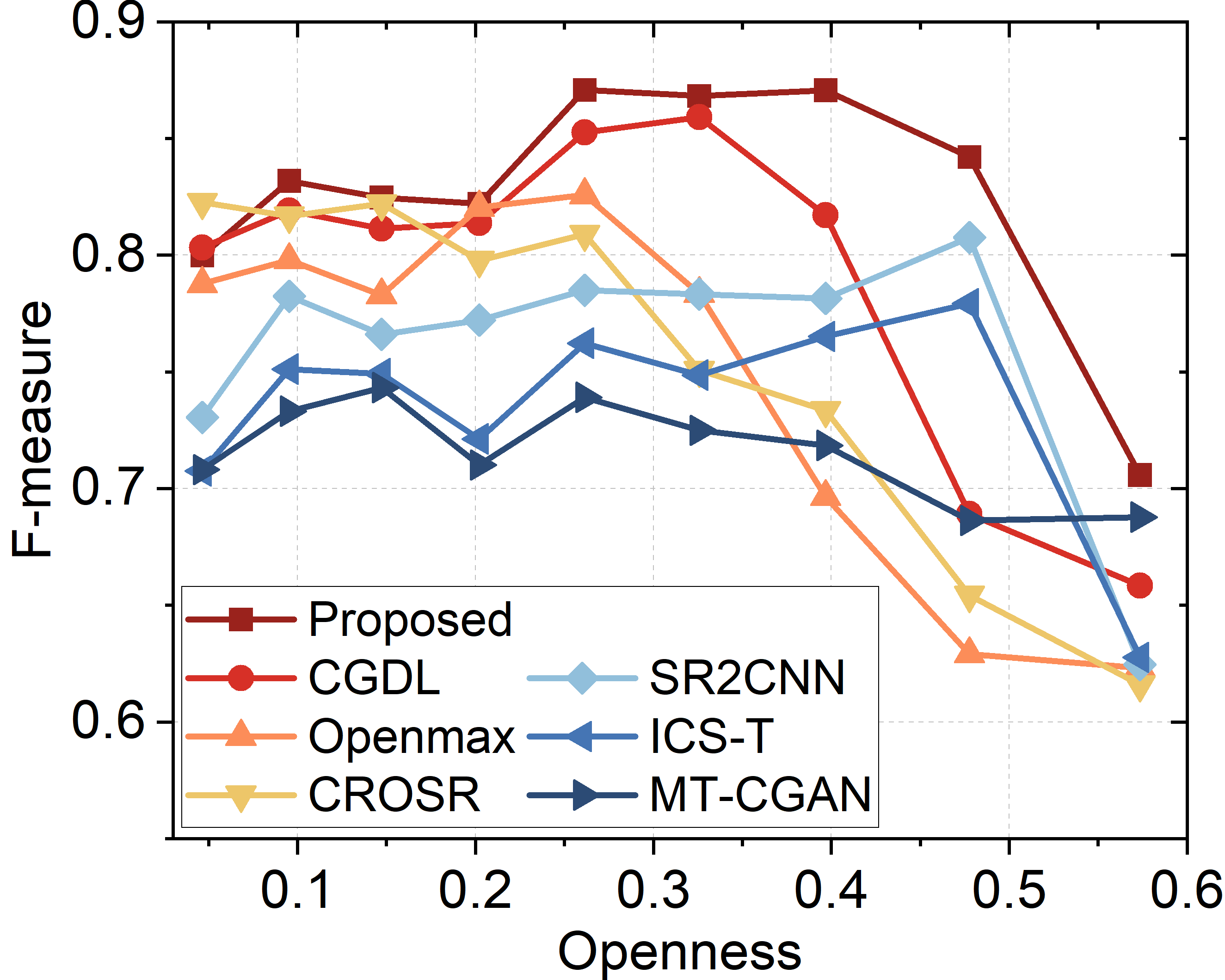}}
      \label{figb4d}
      \captionsetup{justification=justified}
      \caption{The F-measure performance of different approaches against varying Openness.}
      \label{figb4}
  \end{figure}

\par The F-measure of the proposed method and the baseline methods against varying openness are presented in Fig. \ref{figb4}. It can be observed that for SNRs of 4dB, 0dB, and -4 dB, the F-measure of all methods decreases with increasing openness. When openness is above 0.1, the proposed method achieves the best performance across all cases. Moreover, as the openness continues to grow, the margin of improvement in F-measure provided by the proposed method compared to the baseline methods also enlarges, indicating the capability of CIR to handle more complex situations.

\par When the SNR is -8dB, interestingly, the F-measure first increases and then decreases, reaching a peak when the openness is above 0.2 but less than 0.5. This is because F-measure is a metric to evaluate the overall performance. When the openness is low, indicating a lower proportion of unknown categories, the model primarily focuses on the classification task of known categories. Due to the influence of noise and increased feature ambiguity, the model performs poorly in the KCC task, resulting in a decrease in F-measure. When the openness is high, the model performs poorly in the UCI task, leading to a decrease in F-measure. The results show that the proposed method demonstrates a clear performance advantage when openness exceeds 0.3. This observation reflects the stronger capability of our approach in handling extreme scenarios with low SNR and high proportions of unknown classes.


\par
\begin{figure}[!t]
      \centering
      \subfloat[Openness=0.15, SNR=-4dB]{\includegraphics[width=1.72in]{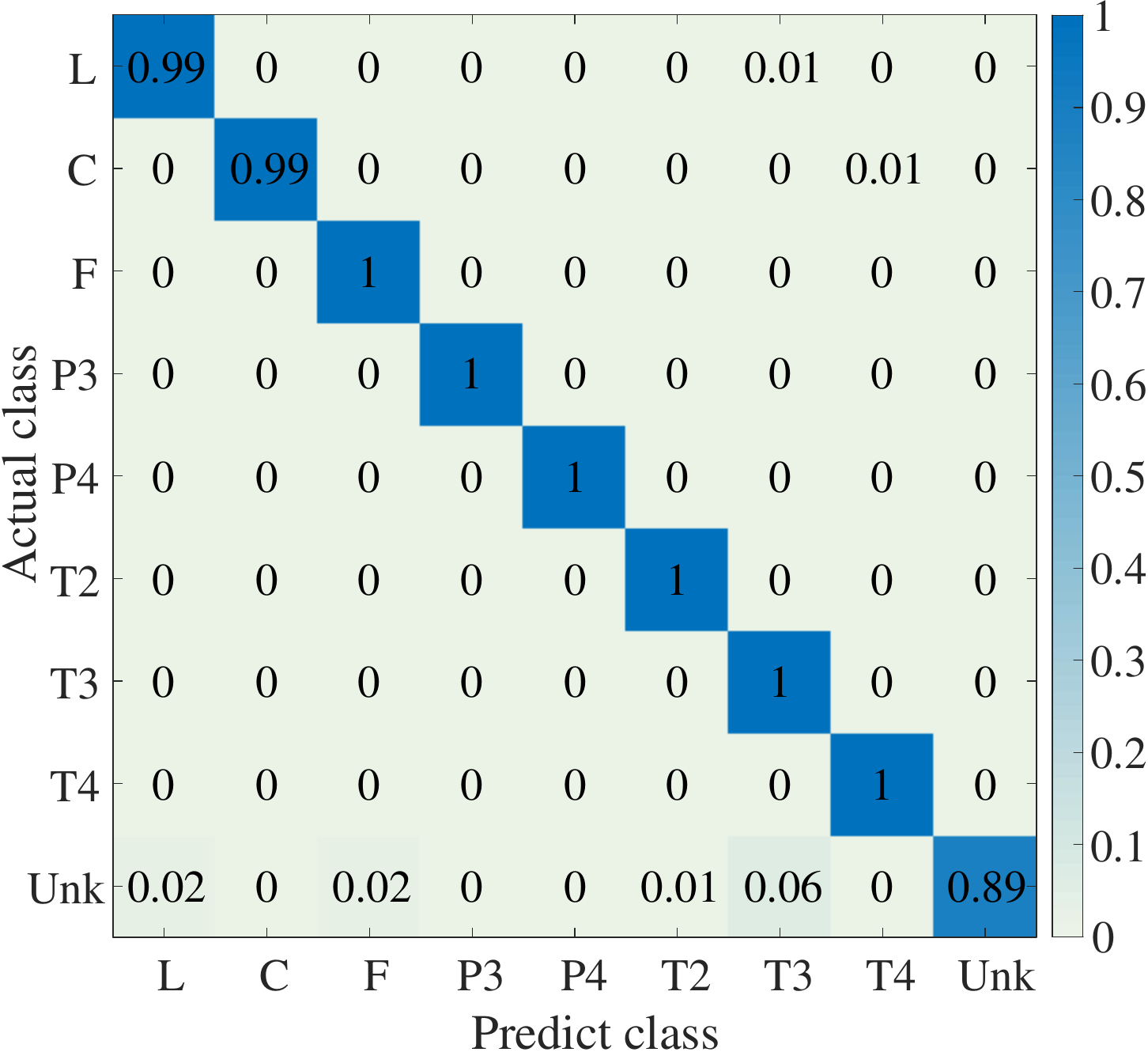}}
      \label{figbca}
      \subfloat[Openness=0.15, SNR=-8dB]{\includegraphics[width=1.72in]{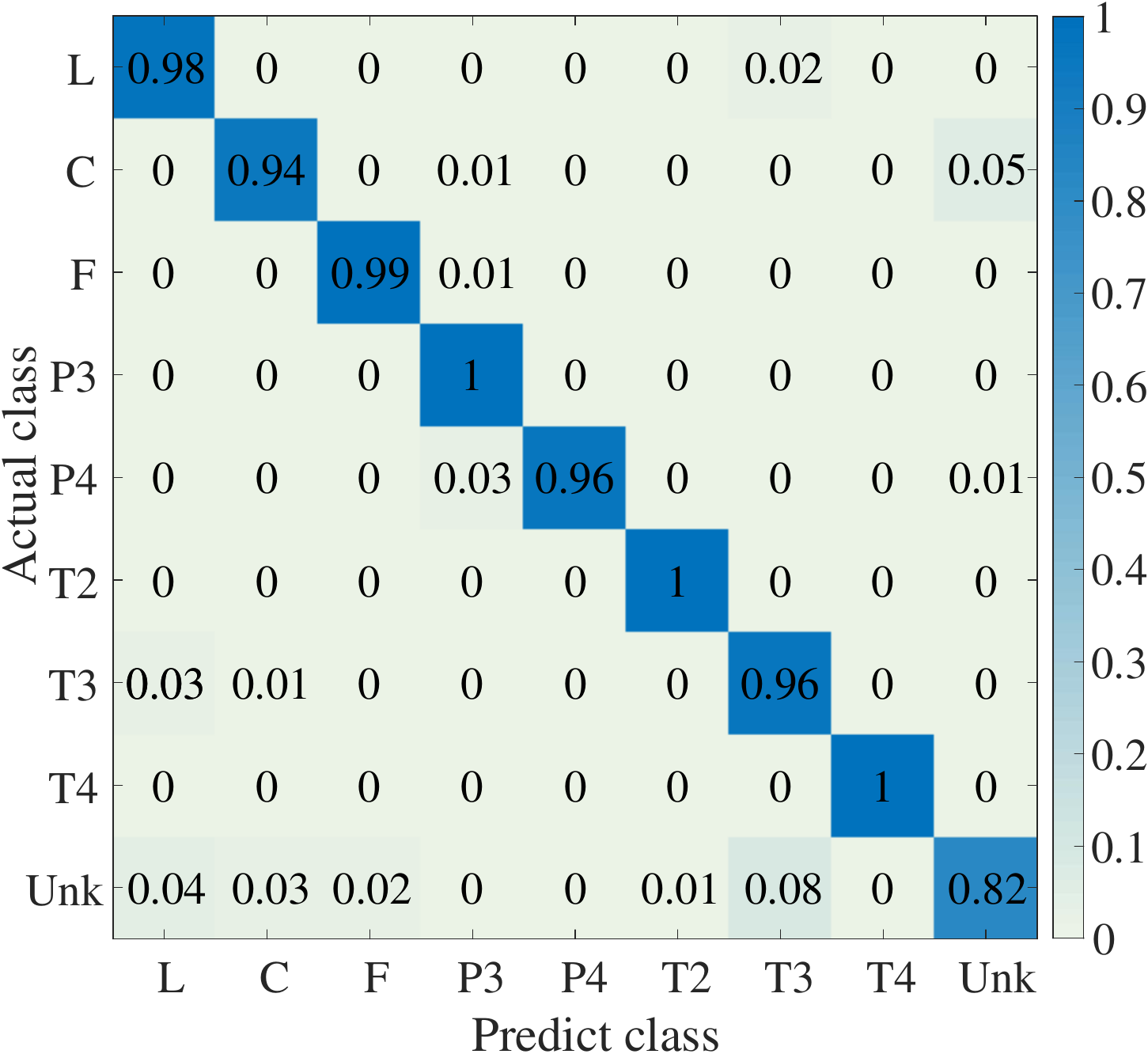}}
      \label{figbcb}
      \subfloat[Openness=0.57, SNR=-4dB]{\includegraphics[width=1.72in]{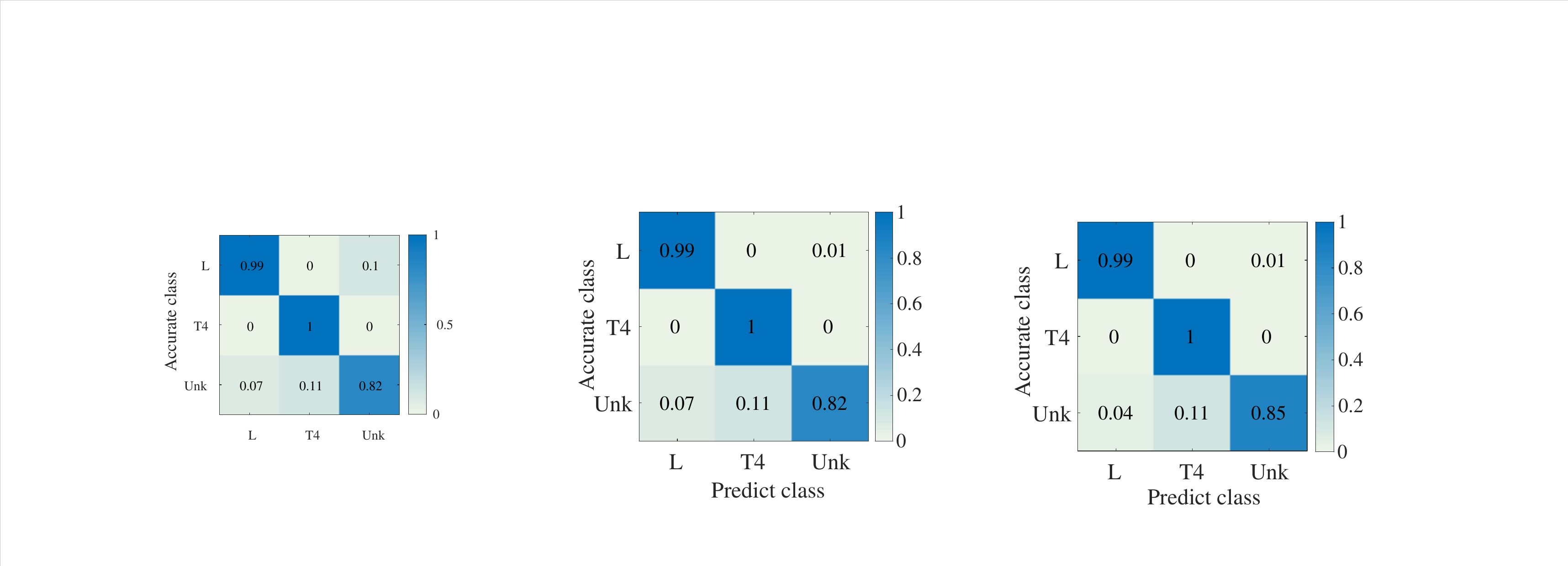}}
      \label{figbcc}
      \subfloat[Openness=0.57, SNR=-8dB]{\includegraphics[width=1.72in]{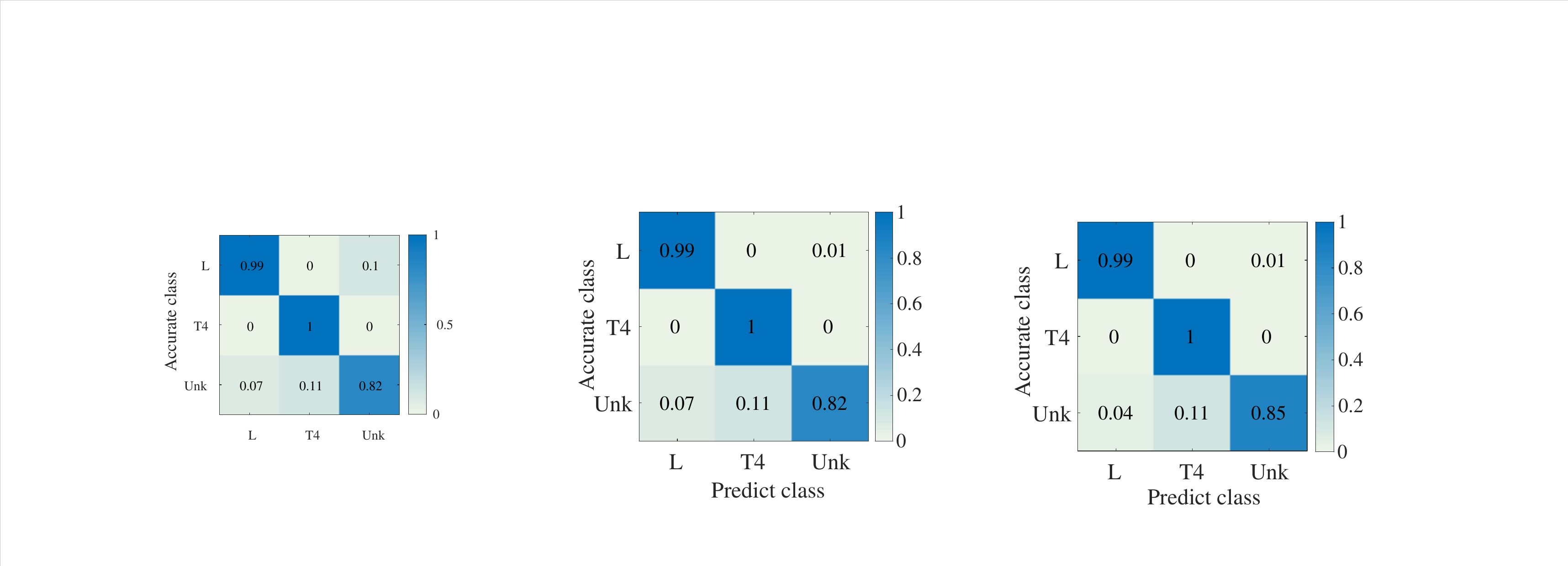}}
      \label{figbcd}
      \captionsetup{justification=justified}
      \caption{The confusion matrices at openness values of 0.15 and 0.57, where the two easily confused modulation type pairs T1\&T3 and T2\&T4 are introduced.}
      \label{figbc}
  \end{figure}

 At openness values of 0.15 and 0.57, there is a noticeable decline in the F-measure. To analyze the underlying causes of the noticeable decline, Fig. \ref{figbc} provides the confusion matrices at -4dB and -8dB. For openness of 0.15, this can be attributed to the introduction of T1 as a new unknown class. The high similarity between the T1 and T3 modulations means that they can easily be confused and misclassified as each other, leading to the sudden drop of UCI performance. Similarly, at openness 0.57, the introduction of T2 as an unknown class results in significant confusion between T2, T4 and LFM, further impacting the performance negatively.

\begin{table*}[!t]
  \centering
  \caption{Comparison of KCC and UCI performance with measured signals}
    \begin{threeparttable}
    \begin{tabular}{p{4.19em}llllllllllllll}
    \toprule
    \multirow{2}[4]{*}{Methods} & \multicolumn{2}{p{8.38em}}{LFM(Unk)} &       & \multicolumn{2}{p{8.38em}}{NLFM (Unk)} &       & \multicolumn{2}{p{8.38em}}{Baker7 (Unk)} &       & \multicolumn{2}{p{8.38em}}{M\&B(Unk)} &       & \multicolumn{2}{p{8.38em}}{L\&N(Unk)} \\
\cmidrule{2-3}\cmidrule{5-6}\cmidrule{8-9}\cmidrule{11-12}\cmidrule{14-15}    \multicolumn{1}{l}{} & \multicolumn{1}{p{2.59em}}{AKS(\%)   } & \multicolumn{1}{p{4.19em}}{AUS(\%)} &       & \multicolumn{1}{p{2.59em}}{AKS(\%)   } & \multicolumn{1}{p{4.19em}}{AUS(\%)} &       & \multicolumn{1}{p{2.59em}}{AKS(\%)   } & \multicolumn{1}{p{4.19em}}{AUS(\%)} &       & \multicolumn{1}{p{2.59em}}{AKS(\%)   } & \multicolumn{1}{p{4.19em}}{AUS(\%)} &       & \multicolumn{1}{p{2.59em}}{AKS(\%)   } & \multicolumn{1}{p{4.19em}}{AUS(\%)} \\
\cmidrule{1-3}\cmidrule{5-6}\cmidrule{8-9}\cmidrule{11-12}\cmidrule{14-15}    Proposed & 100     & 99.4 \textcolor{red}{(+1.6)} &       & 100     & 99.0 \textcolor{red}{(+1.8)}  &       & 99.6 & 99.6 \textcolor{red}{(+6.8)} &       & 100     & 97.7 \textcolor{red}{(+7.4)}&       & 100     & 99.5 \textcolor{red}{(+2.0)} \\
    CGDL  & 99.8 & 97.8 \textcolor{blue}{(-1.6)} &       & 100     & 97.2 \textcolor{blue}{(-1.8)} &       & 99.2 & 90.4 \textcolor{blue}{(-9.2)} &       & 99.6 & 82.3  \textcolor{blue}{(-15.4)}&       & 99.6 & 97.5 \textcolor{blue}{(-2.0)} \\
    CROSR & 99.6 & 96.4 \textcolor{blue}{(-3.0)} &       & 100     & 96.4 \textcolor{blue}{(-2.6)} &       & 100     & 84.4 \textcolor{blue}{(-15.2)} &       & 99.2 & 84.6  \textcolor{blue}{(-13.1)}&       & 100     & 94.3 \textcolor{blue}{(-5.2)} \\
    Openmax & 99.6 & 88.6 \textcolor{blue}{(-10.8)} &       & 99.6 & 90.4 \textcolor{blue}{(-8.6)} &       & 99.6 & 80.4 \textcolor{blue}{(-19.2)} &       & 99.2 & 74.8  \textcolor{blue}{(-22.9)}&       & 100     & 92.5 \textcolor{blue}{(-7.0)} \\
    SR2CNN & 100     & 96.6 \textcolor{blue}{(-2.8)} &       & 100     & 96.2 \textcolor{blue}{(-2.8)} &       & 99.6 & 92.8 \textcolor{blue}{(-6.8)} &       & 99.6 & 90.3  \textcolor{blue}{(-7.4)}&       & 99.6 & 96.8 \textcolor{blue}{(-2.7)} \\
    ICS-T & 99.6 & 90.0 \textcolor{blue}{(-9.4)}  &       & 100     & 86.8 \textcolor{blue}{(-12.2)} &       & 100     & 80.0 \textcolor{blue}{(-19.6)} &       & 100     & 75.0 \textcolor{blue}{(-22.7)} &       & 99.2 & 88.1 \textcolor{blue}{(-11.4)} \\
    MTCGAN & 99.6 & 92.4 \textcolor{blue}{(-7.0)} &       & 100     & 90.0 \textcolor{blue}{(-9.0)} &       & 100     & 84.4 \textcolor{blue}{(-15.2)} &       & 99.6 & 76.6 \textcolor{blue}{(-21.1)} &       & 99.6 & 87.4 \textcolor{blue}{(-12.1)} \\
   \bottomrule
    \end{tabular}%
 \begin{tablenotes}
        \footnotesize
        \item \textcolor{red}{Red text} indicates the pp increase achieved by CIR compared to the best-performing baseline. 
        \item \textcolor{blue}{Blue text} represents the pp decrease for baselines compared to CIR. 
      \end{tablenotes}
 \end{threeparttable}

  \label{label_measure}%
\end{table*}%


\subsubsection{Performance Validation with Measured Signals}

\par The ASK and AUS of all methods with the measured signal dataset D3 are presented in Label. \ref{label_measure}. The AKS of all methods are above 99\% for the five sub-datasets. This can be attributed to the generally high SNR, which leads to very clear separation between known classes in the representation space. For UCI task, the proposed method, SR2CNN and CGDL exhibit relatively better and more stable AUS performance against other methods. Among them, the proposed method achieves the best UCI performance across all five subsets. 

\par Notably, the proposed method exhibits pronounced advantages in specific dataset configurations. When considering Baker7 as the unknown class, the proposed method outperforms the best-performing baseline by 6.8 pp. An even larger margin of 7.4 pp is achieved when both M13 and Baker7 are treated as unknown classes. This improvement can be attributed to the complex modulation patterns exhibited by Baker7 and M15. Overall, the robust performance of the proposed CIR, validates its effectiveness in addressing real-world OSR tasks, particularly in handling more intricate patterns.

\section{Conclusion}
\label{section6}
\par The paper analyzes two challenging issues in AMOSR: sparse feature distribution and inappropriate decision boundary. To tackle these issues, we propose a class information guided reconstruction framework to perform AMOSR task based on reconstruction losses. The CIR fully leverages class information to guide the reconstruction procedure, which facilitates amplifying the distance between discriminative features of the known and unknwon samples, thereby alleviating the adverse impacts incurred by decision boundary selection. To mine the class information, we design class conditional vectors and mutual information loss function. By this design, the CIR attains optimal UCI performance without compromising KCC accuracy, especially in scenarios with a higher proportion of unknown classes and low SNRs. Experiments on simulated and measured samples against the state-of-the-art OSR methods in CV and AMR domains validate the effectiveness and robustness of the proposed CIR.

\par There are several directions that need to be investigated in the future. Firstly, OSR of fingerprint information should be further explored, which is conducive to SEI and detection of illegal user intrusion. Secondly, most AMOSR methods are performed based on TFIs, whereas the high computational complexity for TFA limits the real-time processing ability of these methods. Therefore, investigating efficient sequence-based OSR methods is necessary for practical applications.

\appendices
\section{Proof of Theorems}
\label{app1}
\par Proof of \textbf{Theorem 1.}:
\begin{proof}\renewcommand{\qedsymbol}{}
\begin{equation}
\begin{split}
& \qquad  \mathcal{L}_{UB} - \text{MI} (\bm{Z}, \bm{X}) \\
&= \iint p(z,x) \text{log} p(x|z) \text{d}z\text{d}x - \iint p(z) p(x) \text{log} p(x|z) \text{d}x\text{d}z \\
& \quad - \iint p(z,x)\text{log}\left(\frac{p(z,x)}{p(z)p(x)}\right)\text{d}z\text{d}x  \\
&= \iint p(z,x) \text{log} p(x) \text{d}z\text{d}x - \iint p(z)  p(x) \text{log} p(x|z) \text{d}z\text{d}x \\
&= \int p(x) \text{log} p(x) \text{d}x - \iint p(x)  p(z) \text{log} p(x|z) \text{d}x\text{d}z \\
&= \int p(x) \left( \text{log} p(x)- \int p(z) \text{log} p(x|z) \text{d}z \right)\text{d}x \\
\end{split}
\end{equation}


By definition, the probability distribution of inputs $x$ can be expressed as $p(x) = \int p(x|z)p(z)\text{d}z$.

Taking the logarithm of both sides yields $\text{log} p(x) = \text{log} \int p(x|z)p(z)\text{d}z$. Applying Jensen's inequality to the expectation of the logarithm results in:

\begin{equation}
    \text{log} \int p(x|z)p(z)\text{d}z \geq  \int \text{log}p(x|z) p(z)\text{d}z
\end{equation}

\par Then, the derivation yields the following:

\begin{equation}
\begin{split}
& \qquad  \mathcal{L}_{UB} - \text{MI} (\bm{Z}, \bm{X}) \\
&= \int p(x) \text{d}x \left( \text{log} p(x)- \int p(z) \text{log} p(x|z) \text{d}z \right) \\
&\geq \int p(x) \text{d}x \left( \int \text{log}p(x|z) p(z)\text{d}z- \int p(z) \text{log} p(x|z) \text{d}z \right) = 0
\end{split}
\end{equation}
\par Thus, the upper bound of $MI (\bm{Z}, \bm{X})$ is derived. Since $p(x|z)$ is not accessible, A variational approximation distribution $q_{\theta}(x|z)$ is utilized to approximate it.
\end{proof}

\par Proof of \textbf{Theorem 2.}:
\begin{proof}\renewcommand{\qedsymbol}{}
    \par By definition of MI, the following relationship holds:
    \begin{equation}
        \text{MI}(\bm{Z}, \bm{Y}) = \mathcal{D}_{KL}\left(P(\bm{Z},\bm{Y})||P(\bm{Y})P(\bm{Z})\right)
    \end{equation}
    \par According to the definition of conjugate function estimated by f-divergence, the derivation yields:
    
    \begin{equation}
    \begin{split}
        & \qquad \mathcal{D}_{f}\left(P(\bm{Z},\bm{Y})||P(\bm{Z})P(\bm{Y})\right)\\
        & = \iint p(y)p(z) f\left( \frac{p(y,z)}{p(y)p(z)} \right)\text{d}y\text{d}z\\
        & = \iint p(y)p(z) \left( \mathop{sup}\limits_{{f^{'} \in dom(f^*)}} \left\{ \frac{p(y,z)}{p(y)p(z)}f^{'}  - f^{*}(f^{'} ) \right\}\right)\text{d}y\text{d}z \\
        & \geq \iint p(y)p(z) \left( \frac{p(y,z)}{p(y)p(z)}F(y,z) - f^{*}(F(y,z)) \right)\text{d}y\text{d}z \\
        & = \iint p(y,z) F(y,z)\text{d}y\text{d}z - \iint p(y)p(z)f^{*}(F(y,z))\text{d}y\text{d}z
    \end{split}
    \end{equation}
    \par Then, the KL-divergence can be estimated using $f(y)=y\text{log}y$, deriving the following:
    \begin{equation}
        \begin{split}
            & \qquad \mathcal{D}_{KL}\left(P(\bm{Z},\bm{Y})||P(\bm{Z})P(\bm{Y})\right)\\
            & = \mathop{sup}\limits_{F: \mathbb{Y}\times\mathbb{Z} \rightarrow \mathbb{R}} \mathbb{E}_{p(y,z)}[F(y,z)] - \text{log} \left(\mathbb{E}_{p(y)p(z)}[e^{F(y,z)}] \right)  \\
            & \geq \mathop{sup}\limits_{F_{\phi}} \mathbb{E}_{p(y,z)}[F_{\phi}(y,z)] - \text{log} \left(\mathbb{E}_{p(y)p(z)}[e^{F_{\phi}(y,z)}] \right)  \\
        \end{split}
    \end{equation}
    Thus, the lower bound of $\text{MI} (\bm{Z}, \bm{Y})$ is derived.
\end{proof}

\section{Parameter Settings}
\label{app2}
\begin{table}[h]
  \centering
  \caption{Parameter settings of eleven modulation types}
    \setlength{\tabcolsep}{5mm}{
    \begin{tabular}{lll}
    \toprule
    \multicolumn{1}{l}{Types} & \multicolumn{1}{l}{Parameters} & \multicolumn{1}{l}{Value} \\
    \midrule
    All   &   Carrier frequency    & $U(\frac{1}{6},\frac{1}{4})f_s$ \\
    \midrule
    LFM   &   Bandwidth    & $U(\frac{1}{20},\frac{1}{10})f_s$ \\
    \midrule
    \multirow{2}[2]{*}{Costas} &   Number of frequency hops    & $U(3,6)$ \\
          &   Fundamental frequency    & $U(\frac{1}{32},\frac{1}{20})f_s$ \\
    \midrule
    \multirow{2}[2]{*}{Frank} &   Cycles per phase code    & $U(4,6)$ \\
          &   Number of frequency steps   & $U(4,6)$ \\
    \midrule
    \multirow{2}[2]{*}{P1, P2} &   Cycles per phase code    & $U(4,6)$ \\
          &   Number of frequency steps    & $U(4,6)$ \\
    \midrule
    \multirow{2}[2]{*}{P3, P4} &   Cycles per phase code    & $U(4,6)$ \\
          &   Number of sub-codes    & $\{ 16,25,36 \}$ \\
    \midrule
    \multirow{2}[2]{*}{T1, T2} &   Number of phase states    & $U(2,3)$ \\
          &   Number of segments    & $U(4,6)$ \\
    \midrule
    \multirow{2}[2]{*}{T3, T4} &   Number of phase states    & $U(2,3)$ \\
          &   Bandwidth    & $U(\frac{1}{20},\frac{1}{10})f_s$ \\
    \bottomrule 
    \end{tabular}}%

     \begin{tablenotes}
        \footnotesize
        \item[*] U(a,b) represents uniform sampling.  
      \end{tablenotes}

  \label{table5}%
\end{table}%

\section*{Acknowledgments}
The authors appreciate the editors and anonymous referees for their efforts and constructive comments to improve the quality of this paper.

\bibliographystyle{IEEEtran}
\bibliography{IEEEabrv,rnBIB}
\end{document}